\documentclass[aps,prb,twocolumn,superscriptaddress,amsmath,amssymb,showpacs]{revtex4}
\usepackage{psfrag}
\usepackage{bm}
\usepackage{graphicx}
\usepackage{subfigure}
\usepackage{dcolumn}

\newcommand{\vk}{\vec{k}}
\newcommand{\e}{\varepsilon}

\begin{document}
\title{Excitonic effects in two-dimensional massless Dirac fermions}
\author{Jianhui Wang}
\affiliation{Department of Physics, Indiana University, Bloomington, IN 47405}
\author{H.A. Fertig}
\affiliation{Department of Physics, Indiana University, Bloomington, IN 47405}
\author{Ganpathy Murthy}
\affiliation{Department of Physics and Astronomy, University of Kentucky, Lexington, KY 40506-0055}
\author{L. Brey}
\affiliation{Instituto de Ciencia de Materiales de Madrid (CSIC), Catoblanco, 28049 Madrid, Spain}

\date{\today}

\begin{abstract}
We study excitonic effects in two-dimensional massless Dirac fermions
with Coulomb interactions by solving the ladder approximation to the
Bethe-Salpeter equation. It is found that the general 4-leg vertex
has a power law behavior with the exponent going from real to
complex as the coupling constant is increased. This change of
behavior is manifested in the antisymmetric response, which
displays power law behavior at small wavevectors reminiscent of a critical state,
and a change in this power law from real to complex that is accompanied by
poles in the response function for finite size systems,
suggesting a phase transition
for strong enough interactions.
The density-density response is also calculated, for which no critical
behavior is found.  We demonstrate that exciton correlations
enhance the cusp in the irreducible
polarizability at $2k_F$, leading to a strong increase in the
amplitude of Friedel oscillations
around a charged impurity.
\end{abstract}

\maketitle

\section{\label{sec:into}INTRODUCTION}
Graphene is a two-dimensional honeycomb lattice of carbon atoms.  Near zero doping,
the low energy quasiparticle states are well-described by
two-dimensional massless Dirac fermions (MDF's).
Graphene supports two species of these, centered
at the two inequivalent corners of the Brillouin zone.
At low or zero doping, the properties of graphene are in many
ways quite different than those of a doped semiconductor,
in large part because of the unusual properties of MDF's
\cite{Castro_Neto_RMP,DasSarma2010}.

One interesting class of questions about graphene involve Coulomb interactions.
In a seminal study, Gonzalez, Guinea and Vozmediano \cite{Gonzalez_1994} demonstrated
that weak Coulomb interactions in undoped graphene are marginally irrelevant,
invalidating the basic premise of strong screening that underlies
the standard treatment of an electron gas as a weakly interacting Fermi liquid.
Moreover, simple estimates of the strength of Coulomb interactions, characterized
by an effective fine structure constant $\beta=e^2/\epsilon \hbar v_F$, where
$v_F$ is the speed of electrons near a Dirac point and $\epsilon$ is the
effective dielectric constant due to a substrate upon which the graphene
may be adsorbed, suggest that Coulomb interactions are effectively \textit{large}
($\beta > 1$) if $\epsilon \sim 1$.  This suggests that properties
of graphene near zero doping should be rather different than those
of non-interacting MDF's; for example, a gap may open in the quasiparticle
spectrum \cite{drut123}, so that rather than behaving as a metal the system
would be insulating.  In real experiments there is little evidence for
such dramatic effects of interactions, most likely because disorder
effects overwhelm those of Coulomb interactions \cite{DasSarma2010}.
If so, one may suppose that interaction effects could become
apparent if sufficiently clean graphene samples can be created.
In this work, we study this theoretical clean limit, and focus
on unusual properties which can emerge in linear response functions
for two-dimensional MDF's due to Coulomb interactions.  As we shall
see, these have important consequences for the induced charge distribution
around an impurity, and suggest that the MDF description breaks down
even before a gap opens in the spectrum.

The unusual effects of a Coulomb potential for MDF's are already
apparent in their response to a Coulomb charge $Ze$,
even when there are no interactions among the electrons themselves.
The wavefunctions for this problem are essentially exactly
calculable \cite{shytov2007,pereira166802,biswas205122,terekhov076803,pereira085101,kotov075433}.
One finds that the
$m$-th circular component of an electron wavefunction
has the short distance form
$\psi_m(r) \sim r^{\sqrt{(m+1/2)^2-Z^2\beta^2}-1/2}$ at a distance $r$ from the impurity.
This is an unusual situation in that the exponent is a function of the impurity charge,
so that the wavefunctions have a non-analytic dependence on the potential strength
at short distances.  This cannot be reproduced at any finite order in perturbation
theory.
When $Z\beta$ is above a critical value, the short distance exponent becomes complex,
corresponding to a maximal penetration of the centrifugal barrier
(the effective potential due to
the angular momentum in the radial equation for the wavefunction)
by the electrons.  We refer to this phenomenon
as ``Coulomb implosion'', and it is analogous \cite{shytov2007} to the
breakdown of the vacuum in
the vicinity of a highly charged nucleus
in QED \cite{schwabl_book}. In graphene, this breakdown is accompanied by the
formation of a charge cloud around the impurity with density falling
off as $1/r^2$, which is absent for $Z\beta$ below its critical value.  The
propagation of a short distance effect (penetration of the centrifugal barrier)
to long distances (appearance of the charge cloud) is one of the special properties
of MDF's, and it reflects the absence of any length scale in the Dirac equation
itself.  As we shall see, there are many-body analogs of these phenomenon
which become apparent in some of the linear response functions.

To address the many-body problem, it is preferable to assess the non-analytic
content of a linear response function as a function of momentum rather
than position.  To see how this might be done, we revisit the problem
of non-interacting MDF's in the presence of a Coulomb impurity, and
analyze how the short distance behavior described above is manifested
in a momentum representation.  Not surprisingly, we find that the
scattering wavefunctions, when expressed in a momentum representation,
display power law behavior at large momentum, with an exponent that
changes from real to complex for $Z\beta$ exceeding the same critical
value found in the real space analysis.  The momentum space analysis
in terms of scattering states naturally suggests that one might find
similar behavior in vertex functions evaluated in the ladder
approximation \cite{fetter}.
Ladder diagrams play an important role in interacting systems
because they allow one to incorporate excitonic correlations
in virtual particle-hole pairs that are generated by the interaction.
We analyze this approximation for
a generic four-point vertex function of MDF's with Coulomb interactions,
and find both the non-analytic power law behavior at large momentum
and a change from real to complex exponent, in this case when
$\beta$ exceeds some critical value.  The analysis suggests the
system undergoes a quantum phase transition at this critical value,
since the exponent necessarily behaves in a non-analytic
way as a function of the parameter $\beta$.
Interestingly, we shall see that the equations for
the vertex function suggest that this transition is
infinite order, suggesting the transition may be in
the same universality class as {\it classical} two
dimensional systems undergoing a Kosterlitz-Thouless (KT)
transition \cite{nelson_2002}.

To relate the four point vertex function to measurable quantities,
we use it to form three-point vertex functions which can then be
used to directly compute linear response functions.  Two are of
particular interest.  The density response function
(equivalently, the density-density correlation function) expresses
the screening response of the system to an external potential,
including due to impurities or inhomogeneities in the system.
We find that the non-analyticity of the four-leg vertex does
{\it not} present itself in this particular quantity.  Nevertheless,
we demonstrate that exciton effects, as expressed in the ladder
diagrams, have important quantitative effects for
for doped systems, where we find a strong enhancement of
Friedel oscillations around a charged impurity relative to
the non-interacting case.

The other important response function involves the charge imbalance
between the two sublattices, the antisymmetric response function.
(Some results on this were reported by us previously \cite{asym}.)
Here we indeed find non-analytic behavior analogous to the
Coulomb impurity problem for non-interacting MDF's.
Although this non-analyticity is apparent at large wavevectors
in the four-leg vertex,
the absence
of length scale in the problem leads to power law behavior
emerging at {\it small} wavevectors in the response function.
In particular this applies to an impurity placed asymmetrically
with respect to the graphene sublattices, which is known to induce
very different charge responses on them
\cite{cheianov_2006,Brey_2007b,wehling_2007,yazyev_2007}.
The sublattice-antisymmetric component of this
acquires a power law tail, with exponent that changes from
real to complex above a critical value of $\beta$.
Interestingly, when a finite size cutoff is included
in the calculation, we find poles in the response function,
suggesting a quantum phase transition to a state with
different charge densities on the sublattices, and hence
a state with spontaneously broken chiral symmetry \cite{Khveshchenko,Gorbar2002}.
These poles however merge together into a branch cut in
the thermodynamic limit \cite{asym}, suggesting a state fluctuating
among ones with the chiral symmetry broken in different possible
ways, and no mean-field mass gap.  This may represent a phase
transition that is a precursor to one in which a real gap
develops in the spectrum \cite{drut123}.

This article is organized as follows.
We begin with a study of the single impurity problem in the momentum representation
in Section~\ref{sec:pRep} and identify the signatures
for the short-distance power law behavior and
Coulomb implosion.  With this simpler example to guide us,
in Section~\ref{sec:4leg} we study the ladder
approximation to the Bethe-Salpeter equation for the 4-leg vertex.
This approximation is the many-body analog of the Lippmann-Schwinger
equation studied in Section~\ref{sec:pRep}. We show that the vertex
function has a power law behavior in momentum, and the exponent
changes from real to complex above a certain value of coupling
constant $\beta$.
The interesting properties of the antisymmetric response are
discussed in Section~\ref{anti-res}.  Finally, in
Section~\ref{sec-den} we focus on the density-density
response function in the ladder approximation, and demonstrate
the important quantitative effects caused by excitonic correlations.

\section{\label{sec:pRep} Impurity problem in the momentum representation}
We begin by discussing the problem of MDF's in the presence of
a Coulomb impurity in terms of scattering states.
The standard (Lippmann-Schwinger) equation for scattering states \cite{merzbacher1970}
takes the form
\begin{equation}
\psi^{(+)}(\vec{x})=\psi^{(0)}(\vec{x})+\int d^2\vec{y} G^{\text{DE}}(\vec{x}-\vec{y}) V(\vec{y}) \psi^{(+)}(\vec{y}),\label{psi_r-space}%"-" on the RHS is replaced by "+" in accordance with the correction of the sign in V(r)
\end{equation}
where $G^{\text{DE}}$ is the (matrix) Green's function determined by the differential equation (DE) $(-\hbar v_F \vec{\sigma}\cdot\hat{p}+\varepsilon \openone) G^{\text{DE}}(\vec{x})=\delta^{(2)}(\vec{x})$,
$V(\vec{y})=Ze^2/\epsilon |\vec{y}|$ is the impurity potential
and $\psi^{(0)}$ is an  eigenstate for $V=0$.
Fourier transforming Eq.~(\ref{psi_r-space}), we have
\begin{equation}
\psi^{(+)}(\vec{p})=\psi^{(0)}(\vec{p})+G^{\text{DE}}(\vec{p})\int\frac{d^2p'}{(2\pi)^2}V(\vec{p}-\vec{p}\,^{\prime})\psi^{(+)}(\vec{p}\,^{\prime}).\label{psi_k-space}
\end{equation}
It is convenient to express the Lippmann-Schwinger equation in terms of angular momentum states.
Introducing the angular components
\begin{align}
\psi_m(p)&=\int^{2\pi}_0\frac{d\theta_p}{2\pi}e^{-i m \theta_p}\psi(\vec{p}),\\
%V(|\vec{p}-\vec{p}\,^{\prime}|)=-Ze^2\sum_n e^{-i n (\theta_{\vec{p}}-\theta_{\vec{p}\,^{\prime}})}U_n(p,p'),\\
%U_n(p,p')=f_n(p/p')/p'
V(|\vec{p}-\vec{p}\,^{\prime}|)&=-Ze^2\sum_n e^{-i n (\theta_{\vec{p}}-\theta_{\vec{p}\,^{\prime}})}f_n(p/p')/p',\\
f_n(x)&=\int^{2\pi}_0\frac{d\theta}{2\pi}\frac{e^{-in\theta}}{[1+x^2-2x\cos(\theta)]^{1/2}},
\end{align}
where $\theta_{\vec{p}}$ is the angle between the vector $\vec{p}$ and the $\hat{x}$ axis,
and using
\begin{equation}
G^{\text{DE}}(\vec{p})=\frac{1}{\varepsilon^2-(\hbar v_Fp^2)}[\varepsilon \openone+\hbar v_F (p_x \sigma^x+p_y \sigma^y)],
\end{equation}
where $\sigma^x$ and $\sigma^y$ are the Pauli matrices, Eq.~(\ref{psi_k-space}) may be
written in the form
\begin{eqnarray}
\label{psi1m}
\psi^{(+)}_{1,m}&=&\psi^{(0)}_{1,m}(p)\\ \nonumber
& &-\frac{Ze^2p}{\varepsilon^2-(\hbar v_Fp)^2}[\varepsilon\int^\infty_0 xf_{-m}(x)\psi^{(+)}_{1,m}(xp)dx\\ \nonumber
&&+\hbar v_Fp\int^\infty_0xf_{-(m+1)}(x)\psi^{(+)}_{2,m+1}(xp)dx],\\
\label{psi2mplus1}
\psi^{(+)}_{2,m+1}&=&\psi^{(0)}_{2,m+1}(p)\\ \nonumber
&&-\frac{Ze^2p}{\varepsilon^2-(\hbar v_Fp)^2}[\hbar v_Fp\int^\infty_0 xf_{-m}(x)\psi^{(+)}_{1,m}(xp)dx \\ \nonumber
&&+\varepsilon\int^\infty_0xf_{-(m+1)}(x)\psi^{(+)}_{2,m+1}(xp)dx].
\end{eqnarray}
Motivated by the observation that power law behavior emerges
in the wavefunctions at small distances, we search for power
law solutions at large wavevector.
Using the ansatz
\begin{equation}
\psi^{(+)}_{\alpha,m}(p)=\frac{c_{\alpha,m}}{p^s} \mbox{ for $p\rightarrow\infty$},
\end{equation}
and neglecting terms of lower order of $p$, Eqs.~\ref{psi1m} may be written as
%\begin{eqnarray}
%c_{1,m}-Z\beta I_{m+1}(s)c_{2,m+1}&=&0\nonumber\\
%Z\beta I_m(s)c_{1,m}-c_{2,m+1}&=&0,
%\label{carray}
%\end{eqnarray}
\begin{eqnarray}
\left(
\begin{array}{cc}
1 & -Z\beta I_{m+1}(s) \\
Z\beta I_m(s) & -1 \\
\end{array}
\right)
\left(
\begin{array}{c}
c_{1,m} \\
c_{2,m+1} \\
\end{array}
\right)
=0
\label{carray}
\end{eqnarray}
where
\begin{equation}
I_m(s)=\int^\infty_0x^{1-s}f_{-m}(x)dx.
\end{equation}
Eqs. \ref{carray} will have non-vanishing solutions provided
\begin{equation}
1-(Z\beta)^2I_m(s)I_{m+1}(s)=0.\label{secularEq}
\end{equation}
One may easily show that $I_m(s)$ has a minimum at $s=3/2$ for any integer $m$,
so for $Z\beta$ larger than a critical $(Z\beta)_c$, the solution $s$
to Eq. \ref{secularEq}
becomes complex. For $m=0$,
$(Z\beta)_c=1/2$; for $m=1$, $(Z\beta)_c=3/2$, etc. This change of behavior
corresponds to that found in the
real space analysis of the Coulomb impurity problem
\cite{shytov2007}. For $Z\beta > (Z\beta)_c$
(for a given $m$), the complex values of $s$ cause
$\psi^{(+)}_{\alpha,m}(p)$ to oscillate at small $r$,
with no well-defined value of $\psi^{(+)}_{\alpha,m}(r)$
as $r \rightarrow 0$.  This leads to an ill-defined problem
unless a boundary condition for small but finite $r$ is imposed \cite{shytov2007}.
This suggests that some quantities are sensitive to the short
scale cutoff in the problem, a behavior which we will see
holds true as well for some response response functions
when Coulomb interactions are included.  Using the above
analysis as a guide, we now turn to this more complicated
problem.

\section{\label{sec:4leg} General 4-leg vertex}

As we saw in the last section, the signature of Coulomb implosion
in the momentum representation is that the exponent of the power law
of the wavefunction becomes complex. The basic physics in
Eq.~(\ref{psi_r-space}) is clear: in a
perturbative expansion in $V$,
the electron can be scattered
arbitrarily many times by the impurity, and the non-analytic,
power law behavior emerges from a
superposition of {\it all} these possibilities.
If we substitute the electron-impurity scattering with electron-hole
scattering due to Coulomb interactions, we may expect a many-body analog
of both the power law behavior and of Coulomb implosion.
The signature of these should be contained in the general 4-leg vertex
function in the electron-hole channel.

\begin{figure}
\begin{center}
\label{fig:4leg_diagram}\includegraphics[scale=0.65]{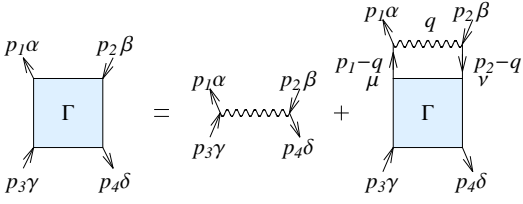}
\end{center}
\caption{\label{4legDiagram} Ladder approximation to the Bethe-Salpeter equation.
}
\end{figure}

To compute this, we note \cite{fetter} that the ladder approximation to
the 4-leg vertex (Fig.~\ref{4legDiagram}) has the same structure
of multiple scattering of an electron from a hole as does
an expansion of Eq. \ref{psi_r-space} in powers of $V$.
The Bethe-Salpeter equation resulting from this ladder sum
has the form \cite{fetter}
\begin{eqnarray}
\lefteqn{\Gamma_{\alpha\beta,\gamma\delta}(p_1,p_2;p_3,p_4)=U(p_1-p_3)\delta_{\alpha\gamma}\delta_{\beta\delta}} \label{BSeq}  \\ \nonumber
&&+\frac{i}{\hbar}\int \frac{d^3q}{(2\pi)^3}U(q)G^{(0)}_{\alpha\mu}(p_1-q)G^{(0)}_{\nu\beta}(p_2-q)\Gamma_{\mu\nu,\gamma\delta}(q),
\end{eqnarray}
where $\Gamma$ is the vertex function whose arguments
are three momenta [spatial components (momentum) $\vec{p}$ and time component (frequency) $p_0$ ],
$U(p)=2\pi e^2/\epsilon|\vec{p}|$ is the Coulomb interaction, and
\begin{eqnarray}
\lefteqn{G^{(0)}(p_0,\vec{p}) =} \label{G0} \\ \nonumber
&& \frac{p_0+\mu/\hbar+v_F \vec{p} \cdot \vec{\sigma}}{(p_0+\mu/\hbar)^2-(v_F \vec{p})^2+i \delta \cdot \text{sgn}(p_0) (p_0+\mu/\hbar)}
\end{eqnarray}
 is the time-ordered Green's function for non-interacting Dirac fermions, and we have allowed the possibility of a non-zero chemical potential $\mu$.

Following Ref. \onlinecite{fetter}, we
introduce two new functions $Q$ and $\chi$ via the relations
\begin{eqnarray}
\lefteqn{\Gamma_{\alpha\beta,\gamma\delta}(p_1,p_2;p_3,p_4)=} \\\nonumber
&&\int\frac{d^3q}{(2\pi)^3}U(q)Q_{\alpha\beta,\gamma\delta}(p_1-q,p_2-q;p_3,p_4),\\
\lefteqn{\chi_{\alpha\beta,\gamma\delta}(\vec{p},\vec{p}\,^{\prime};P)=} \\ \nonumber
&&\int\frac{dp_0}{2\pi}Q_{\alpha\beta,\gamma\delta}(p+\frac{1}{2}P,p-\frac{1}{2}P;p'+\frac{1}{2}P,p'-\frac{1}{2}P).
\end{eqnarray}
In these expressions, $P$ represents the four-momentum of the particle-hole pair, which
may be understood as entering the vertex at the bottom of the
diagrams in Fig. \ref{4legDiagram} and exiting at the top.  In this interpretation, $p$ then represents the
relative momentum of the pair before the collision, and $p^{\prime}$ the momentum
afterwards.  These equations
allow Eq.~(\ref{BSeq}) to take the form
\begin{eqnarray}
\label{eq-chi}
\lefteqn{\chi_{\alpha\beta,\gamma\delta}(\vec{p},\vec{p}\,^{\prime};\vec{P})=\delta_{\alpha\gamma}\delta_{\beta\delta}(2\pi)^2\delta^{(2)}(\vec{p}\,^{\prime}-\vec{p})}\\ \nonumber
&&+K_{\alpha\beta,\mu\nu}(\vec{p},\vec{P})\int\frac{d^2q}{(2\pi)^2}U(\vec{q})\chi_{\mu\nu,\gamma\delta}(\vec{p}-\vec{q},\vec{p}\,^{\prime};\vec{P}),
\end{eqnarray}%if K doesn't depend on P_0, chi shouldn't
where
\begin{equation}
K_{\alpha\beta,\mu\nu}(\vec{p},\vec{P})\equiv\frac{i}{\hbar}\int\frac{dp_0}{2\pi}G^{(0)}_{\alpha\mu}(p+\frac{1}{2}P)G^{(0)}_{\nu\beta}(p-\frac{1}{2}P).
\label{eq-K}
\end{equation}
Note that because we have integrated over $p_0$, there is no frequency dependence
in $K$ or $\chi$ \cite{fetter}.  The quantity $\chi$ can be used to directly to
construct static response functions, as we shall see below.

In the case of the Dirac particle colliding with a Coulomb impurity we found
power law behavior for large momenta, specifically from collisions involving
a large change in momentum ($|\vec{p}|>>|\vec{p}^{\prime}|$ in Eq. \ref{psi_k-space}.)
Since in the two-body problem, the particle
and hole scatter from one another,
we search for analogous behavior in the vertex functions at large momentum
difference $|\vec{p}-\vec{p}\,^{\prime}|$.
For example,
for fixed $P$ and $\vec{p}\,^{\prime}$, when $|\vec{p}\,|\rightarrow \infty$, the equations for $\chi_{11,\gamma\delta}$ and $\chi_{22,\gamma\delta}$ become

\begin{eqnarray}
\label{eq-chi11}
\lefteqn{\chi_{11,\gamma\delta}(\vec{p}\,)=}\\ \nonumber
 &&\frac{1}{4\hbar v_F|\vec{p}\,|}\int\frac{d^2q}{(2\pi)^2}U(\vec{p}-\vec{q})[\chi_{11,\gamma\delta}(\vec{q})-\chi_{22,\gamma\delta}(\vec{q})],\\
\label{eq-chi22}
\lefteqn{\chi_{22,\gamma\delta}(\vec{p}\,)=}\\ \nonumber
 &&\frac{1}{4\hbar v_F|\vec{p}\,|}\int\frac{d^2q}{(2\pi)^2}U(\vec{p}-\vec{q})[-\chi_{11,\gamma\delta}(\vec{q})+\chi_{22,\gamma\delta}(\vec{q})].
\end{eqnarray}
Note in these expressions terms of order $P/p$, $p\,^{\prime}/p$ have been dropped,
and $\vec{p}$ and $P$ are suppressed in the arguments of $\chi$.

Introducing circular moments as before, i.e.%note page 89
\begin{equation}
\chi^{(m)}=\int ^{2\pi}_0 \frac{d\theta_p}{2\pi}e^{-i m \theta_p}\chi(\vec{p}),
\end{equation}
and using the ansatz
\begin{align}
\chi^{(m)}_{11,\gamma\delta}(p)&=\frac{C^{(m)}_{11,\gamma\delta}}{p^s},\\
\chi^{(m)}_{22,\gamma\delta}(p)&=\frac{C^{(m)}_{22,\gamma\delta}}{p^s},
\end{align}
the $m$-th angular component of the above coupled equations reduces to
\begin{align}
 \left(\begin{array}{c c}\frac{4}{\beta}-I_m(s) & I_{m}(s)\\ I_m(s) & \frac{4}{\beta}-I_{m}(s) \end{array}\right)
 \left(\begin{array}{c}C^{(m)}_{11,\gamma\delta} \\ C^{(m)}_{22,\gamma\delta}\end{array}\right) = 0.
\end{align}
Nontrivial solutions to this set of homogeneous linear equations may be found if
\begin{equation}
2/\beta=I_m(s), \label{eq-C-11-22}
\end{equation}
which determines the exponent $s$ for a given $\beta$. The critical $\beta$ where the exponent of the power law becomes complex is $\beta_c=2/I_m(3/2)$.  We thus see the many-body problem has
a Coulomb implosion instability analogous to what is found in the Coulomb impurity
problem for non-interacting MDF's.  A natural interpretation for this instability
is that it indicates a transition to a gapped, excitonic insulator state
\cite{Gorbar2002,Khveshchenko}.
However, as we shall see below, when analyzed in terms of the appropriate
linear response function, such an interpretation is only consistent when
considering a finite size system \cite{asym}; in the thermodynamic limit
the transition is likely to a state with a fluctuating mass gap.

$\chi^{(m)}_{11,\gamma\delta}$ and $\chi^{(m)}_{22,\gamma\delta}$ are not
the only components of the vertex function which display instabilities.
An analogous instability appears
for $\chi_{12,\gamma\delta}$ and $\chi_{21,\gamma\delta}$,
which are governed by the equations
\begin{eqnarray}
\lefteqn{\chi_{12,\gamma\delta}(\vec{p})=}\\ \nonumber
 &&\frac{1}{4\hbar v_F|\vec{p}|}\int\frac{d^2q}{(2\pi)^2}U(\vec{p}-\vec{q})[\chi_{12,\gamma\delta}(\vec{q})-e^{-2i\theta_{\vec{p}}}\chi_{21,\gamma\delta}(\vec{q})],\\
\lefteqn{\chi_{21,\gamma\delta}(\vec{p})=}\\ \nonumber
 && \frac{1}{4\hbar v_F|\vec{p}|}\int\frac{d^2q}{(2\pi)^2}U(\vec{p}-\vec{q})[-e^{2i\theta_{\vec{p}}}\chi_{12,\gamma\delta}(\vec{q})+\chi_{21,\gamma\delta}(\vec{q})].
\end{eqnarray}
The equations for the coefficients in the power law ansatz are
\begin{align}
 \left(\begin{array}{c c}\frac{4}{\beta}-I_m(s) & I_{m+2}(s)\\ I_m(s) & \frac{4}{\beta}-I_{m+2}(s) \end{array}\right)
 \left(\begin{array}{c}C^{(m)}_{12,\gamma\delta} \\ C^{(m+2)}_{21,\gamma\delta}\end{array}\right) = 0,
\end{align}
so that
the equation for $s$ is
\begin{equation}
4/\beta =I_m(s)+I_{m+2}(s).\label{eq-C-12-21}
\end{equation}

The critical $\beta$ for the 12 and 21 components of $\chi$ is $\beta_c'=4/[I_m(3/2)+I_{m+2}(3/2)]>\beta_c$.
This suggests that $\chi^{(m)}_{11,\gamma\delta}$ and $\chi^{(m)}_{22,\gamma\delta}$
can develop complex exponents before
$\chi^{(m)}_{12,\gamma\delta}$ and $\chi^{(m)}_{21,\gamma\delta}$ as
$\beta$ is increased from small values.  We will see more generally
that certain combinations of the vertex functions do not appear
to develop power law behavior at all; most importantly this appears
to be the case for the vertex function relevant to the density-density response function.
In this context, we
note that the coefficients satisfy the conditions $C^{(m)}_{11,\gamma\delta}=-C^{(m)}_{22,\gamma\delta}$ and $C^{(m)}_{12,\gamma\delta}=-C^{(m+2)}_{21,\gamma\delta}$, so that $\chi_{11,\gamma\delta}(\vec{p})=-\chi_{22,\gamma\delta}(\vec{p})$ and $\chi_{12,\gamma\delta}(\vec{p})=-\chi_{21,\gamma\delta}(\vec{p})$.
These relations among the components of  $\chi_{\alpha\beta,\gamma\delta}$
can also be seen from Eqs. \ref{eq-chi} and \ref{eq-K} when expressed in terms
of circular moments.   The density-density response turns out to involve
$\chi_{\alpha \alpha,\gamma \gamma}$  (repeated indices here are summed),
so that the power law behavior is canceled away.

\begin{figure}
\begin{center}
\subfigure[]{\label{fig:beta01}\includegraphics[width=0.75\columnwidth]{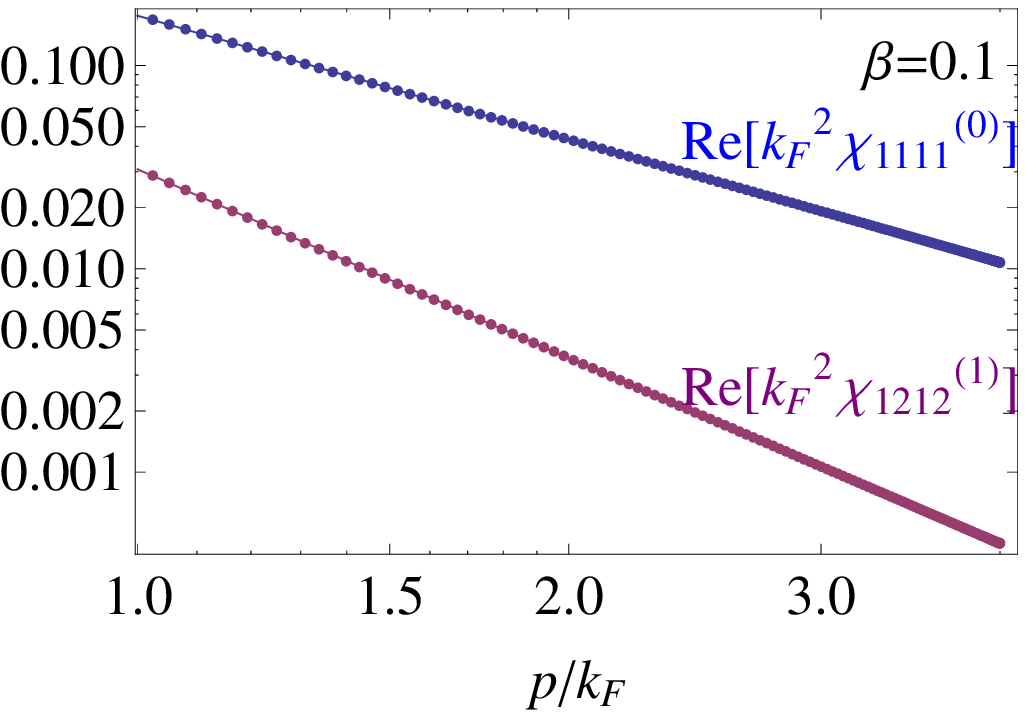}}\\
\subfigure[]{\label{fig:ReChi1111m0}\includegraphics[width=0.75\columnwidth]{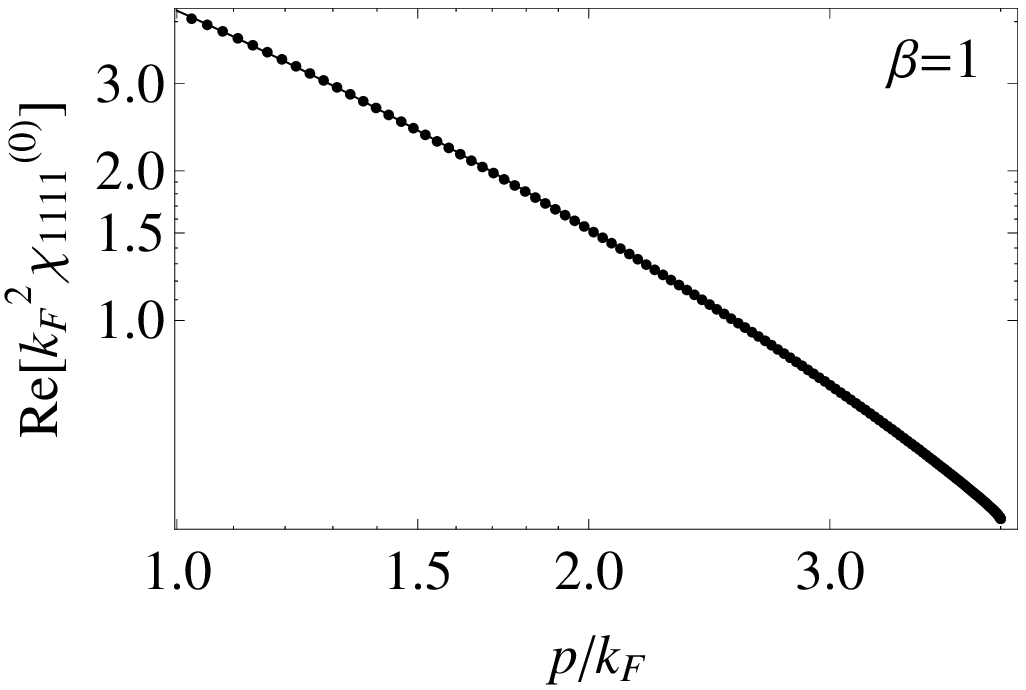}}
\end{center}
\caption{\label{powerlawEg} (Color online) Some examples of power laws from solving Eq.~(\ref{eq-chi}) numerically. $P=0$ and $p\,^{\prime}=k_F/2$. (a) $\beta=0.1<\beta_c$, the solid lines are fits with the model $C/\tilde{p}^s$, where $\tilde{p}\equiv p/k_F$. For $\Re \chi^{(0)}_{1111}$ we get $s=2.02$ from fitting, $s=1.95$ from solving Eq.~(\ref{eq-C-11-22}). For $\Re \chi^{(1)}_{1212}$ we get $s=3.09$ from fitting, $s=3.05$ from solving Eq.~(\ref{eq-C-12-21}). (b) $\beta=1>\beta_c$. The solid line is a fit with $C \cos \left( s''\log \tilde{p} + \delta \right)/\tilde{p}^{s'}$. The fitting gives $s'=1.57$ and $s''=0.78$, while Eq.~(\ref{eq-C-11-22}) gives $s'=1.5$, $s''=0.60$.
}
\end{figure}

We have verified our analysis by numerically
solving the Bethe-Salpeter equation in the form of Eq.~(\ref{eq-chi}).
We use polar coordinates for the integration and change each dimension of the integration to a discrete sum, independent of the other dimension, i.e.
\begin{equation}
\int^{2\pi}_0 d\theta \int^{\Lambda}_0 dq f(\theta,q) \rightarrow \sum_{i=1}^{N_{\theta}} w_i \sum_{j=1}^{N_q} w_j f(\theta _i,q_j),
\end{equation}
where $\Lambda$ is the momentum cutoff and $f$ denotes a general integrand, and the discretization %discrete coordinates $\theta_i,\,i=1,\dots,N_{\theta};\,q_j,\,j=1,\dots,N_q$ and the corresponding weights $w_i$, $w_j$
in each dimension is done according to the Gauss-Legendre rule \cite{NumericalRecipe}. Now the integral equation is changed to a set of linear equations and can be solved using any existing subroutine, e.g. the appropriate routine in the Lapack package. %see chi6GLe in cooperon
Some examples of our results are shown in Fig.~\ref{powerlawEg}.  As can
be seen, the vertex functions follow power law forms, and the exponents
are close to those expected from our asymptotic analysis, both below and
above the critical value of $\beta$.

An interesting aspect of these results is that the change of the exponent $s=s^{\prime}+is^{\prime\prime}$
from real to complex at a finite value of $\beta$
(i.e., $s^{\prime\prime}$ is zero for $\beta<\beta_c$ but is non-vanishing for $\beta>\beta_c$)
suggests that quantities calculated
from it will have a non-analyticity at $\beta_c$.  However, any such singularity
must be of infinite order.
For example,
if there is a cusp in $\chi$ at $\beta=\beta_c$, then there
would be a contribution
of the form
$B_{\alpha \beta,\gamma \delta} (\vec{p},\vec{p}\,^{\prime};P) \delta (\beta-\beta_c)$
in $\partial ^2 \chi / \partial \beta^2$. Differentiating Eq.~(\ref{eq-chi})
with respect to $\beta$ twice, and requiring that the coefficients of
$\delta (\beta-\beta_c)$ on both sides of the equation are the same,
we find an equation for $B$,
\begin{eqnarray}
\lefteqn{B_{\alpha\beta,\gamma\delta}(\vec{p},\vec{p}\,^{\prime};P)=}\label{eq-B}\\ \nonumber
&&+K_{\alpha\beta,\mu\nu}(\vec{p},\vec{P})\int\frac{d^2q}{(2\pi)^2}
\left[U(\vec{q})|_{\beta=\beta_c}\right]B_{\mu\nu,\gamma\delta}(\vec{p}-\vec{q},\vec{p}\,^{\prime};P).
\end{eqnarray}
This
is nothing but a homogeneous version of Eq.~(\ref{eq-chi}),
with $\beta$ [in $U(\vec{q})$] set to $\beta_c$. If Eq.~(\ref{eq-B}) has
a non-vanishing solution, then Eq.~(\ref{eq-chi}) would also have a
contribution from such a solution, and we would expect $\chi$ to be
divergent at $\beta=\beta_c$.  Our explicit solutions, both in the
asymptotic and the numerical analysis, show that this is not the
case, so that no cusp can be present.  Similarly, higher
order derivatives of $\chi$ also cannot have cusps.  It follows that
the singular behavior at $\beta=\beta_c$ -- and any phase transition
it may represent -- is of {\it infinite} order,
a property it shares with KT transitions \cite{nelson_2002}.
In our analysis of the antisymmetric response below we shall see further
hints of a connection to the KT universality class in this system.

%\end{widetext}

The symmetries of the components of $\chi$ discussed above
led to a cancellation of the power law behavior in the
density-density response function.  This same symmetry suggests that in
a response function of the form
$\sigma^z_{\alpha\beta}\chi_{\alpha\beta,\gamma \delta}$,
with $\sigma^z$ the Pauli matrix,
the power law should be retained.
%$\sigma^z_{\alpha\beta}\chi{\alpha\beta,\gamma \delta}\sigma^z_{\gamma \delta}$
This represents the response of the density difference between the $A$ and $B$
sublattices due to a potential that is antisymmetric in sublattice index.
Any potential that breaks sublattice symmetry will have a component of this
antisymmetric response, so that it is in principle physically accessible.
As we shall show next, the antisymmetric response does capture the power
law as well as the Coulomb implosion physics.

\section{\label{anti-res}antisymmetric response}
We define
the antisymmetric response as
\begin{eqnarray}
M(\vec{q})=-\frac{i}{A}\int^\infty_0dt\left<[\hat{m}_z(-\vec{q},t),\hat{m}_z(\vec{q},0)]\right>,\\
\hat{m}_z(\vec{q})=\sigma^z_{\alpha\beta}\hat{\rho}_{\alpha\beta}(\vec{q}),\mbox{       }
\hat{\rho}_{\alpha\beta}\equiv \sum_{\vec{k}}a^\dagger_{\vec{k}+\vec{q},\alpha}a_{\vec{k},\beta},
\label{M(q)}\end{eqnarray}
where $A$ is the area of the sample, repeated indices are summed,
and henceforward we will set $\hbar=1$.
In principle $M(\vec{q})$ can be determined experimentally by measuring the screening charge induced by an impurity placed asymmetrically with respect to the two sublattices, i.e. anywhere except in the center of a hexagon, or at the middle point of a carbon-carbon bond. The difference between densities on the two sublattices $\hat{m}_z$ is an interesting operator because
when $<\hat{m}_z(\vec{q})>$ is non-vanishing, there is a dynamically generated Dirac mass \cite{Khveshchenko,Gorbar2002}.

\begin{figure}
\begin{center}
%\hspace{-5mm}
\subfigure[]{\label{fig:3leg_diagram}\includegraphics[scale=0.45]{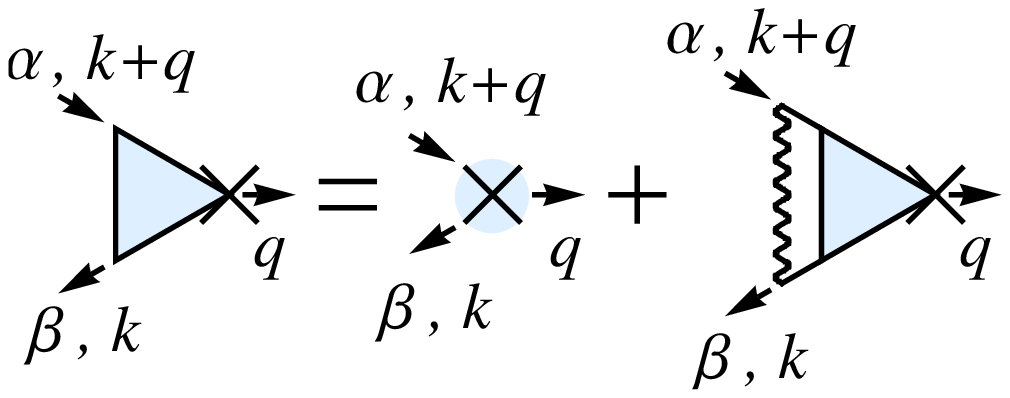}}
%\hspace{-5mm}
\subfigure[]{\label{fig:Mq_diagram}\includegraphics[scale=0.45,trim=2.1cm 0 2.1cm 0,clip]{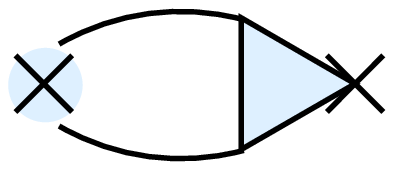}}
\end{center}
\caption{\label{FeynmanDiagrams} (Color online) (a) Diagrammatical equation for the 3-leg vertex $\tilde{\Gamma}^M_{\alpha\beta}(\vec{k},\vec{q})$ with $\sigma^z$ as the zeroth order vertex (the shaded cross in the figure). (b) Diagram for $M(\vec{q})$.
}
\end{figure}

\subsection{Diagrammatic Expansion and Ladder Approximation}

\begin{figure}
\begin{center}
\subfigure[]{\label{fig:reducible}\includegraphics[scale=0.45,trim=2.1cm 0 2.1cm 0,clip]{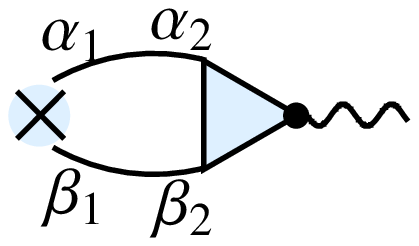}}
\subfigure[]{\label{fig:GamTild}\includegraphics[scale=0.45]{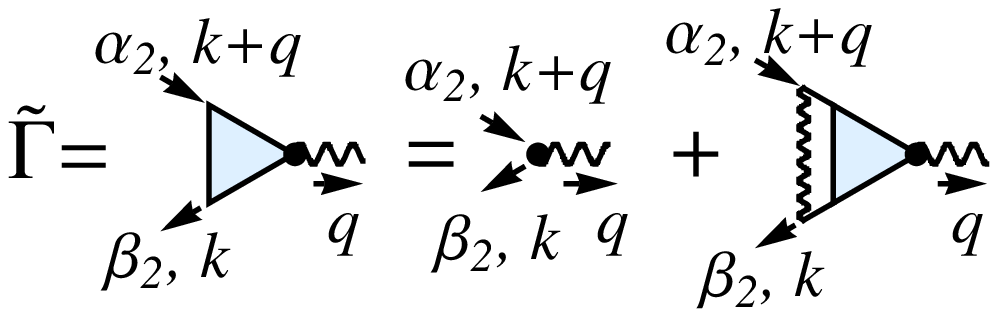}}
\end{center}
\caption{\label{fig-Red-Gam} (Color online) (a) An insertion for the reducible diagram for $M(\vec{q})$. (b) The diagram equation for the 3-leg Coulomb vertex $\tilde{\Gamma}_{\alpha_2 \beta_2}(\vec{k},q)$.}
\end{figure}

The diagrammatic representation for $M(\vec{q})$ is illustrated in Fig. \ref{fig:Mq_diagram},
along with the summation of ladder diagrams, Fig. \ref{fig:3leg_diagram}, representing our approximation for the vertex function.  Notice there are no bubble diagrams
in the diagrammatic expansion of $M(\vec{q})$; there are only
irreducible diagrams.
Reducible diagrams for this quantity turn out to vanish, as we now show.
Any reducible diagram will have at its end an insertion
of the form illustrated in Fig.~\ref{fig:reducible}, in which there
is a Coulomb vertex $\tilde{\Gamma}$.  This insertion represents a multiplicative
contribution to the diagram of the form
\begin{align}
\label{insertion}
& i\int \frac{d^3 k}{(2\pi)^3} \sigma^z_{\alpha_1 \beta_1}G^{(0)}_{\alpha_2\alpha_1}(k+q)G^{(0)}_{\beta_1\beta_2}(k) \tilde{\Gamma}_{\alpha_2 \beta_2}(\vec{k},q) \\ \nonumber
& \equiv \int \frac{d^2k}{(2\pi)^2}\sigma^z_{\alpha_1\beta_1}\tilde{K}_{\alpha_1\beta_1\alpha_2\beta_2}(\vec{k},\vec{q})\tilde{\Gamma}_{\alpha_2 \beta_2}(\vec{k},\vec{q}) \\ \nonumber
& \equiv \int \frac{d^2k}{(2\pi)^2} \sigma^z_{\alpha_1\beta_1}\tilde{\chi}_{\alpha_1 \beta_1}(\vec{k},\vec{q})
\end{align}
In these expressions the overhead tilde denotes quantities pertaining to
3-leg vertex.
Note also that we have set $q_0=0$ in the second line.
Assuming the Coulomb vertex is given by a ladder sum,
the equation determining $\tilde{\chi}$ is [Fig.~\ref{fig:GamTild}]
\begin{eqnarray}
\label{eq-tild-Gam}
\lefteqn{\tilde{\Gamma}_{\alpha_2\beta_2}({\vec k},{\vec q})=\delta_{\alpha_2\beta_2}+}\\ \nonumber
&&\int\frac{d^2q'}{(2\pi)^2}
U(|\vec{q'}|)
\tilde{K}_{\alpha_2\beta_2\gamma_1\gamma_2}(\vec{k}-\vec{q}',\vec{q})\tilde{\Gamma}_{\gamma_1\gamma_2}(\vec{k}-\vec{q}',\vec{q}),
\end{eqnarray}
so that $\tilde{\chi}$ satisfies
\begin{eqnarray}
\label{Coulomb_chi_equation}
\lefteqn{\tilde{\chi}_{\alpha\beta}(\vec{k},\vec{q})=\tilde{K}_{\alpha\beta\alpha_2\alpha_2}(\vec{k},\vec{q})}\\ \nonumber
&&+\tilde{K}_{\alpha\beta\alpha_2\beta_2}(\vec{k},\vec{q})\int\frac{d^2q'}{(2\pi)^2}U(|\vec{q}'|)\tilde{\chi}_{\alpha_2\beta_2}(\vec{k}-\vec{q}',\vec{q}).
\end{eqnarray}

More explicitly, the 11 and 22 components of Eq. \ref{Coulomb_chi_equation}, in the undoped case ($\mu=0$), are
\begin{eqnarray}
\lefteqn{\tilde{\chi}_{11}(\vec{k},\vec{q})=\frac{1-e^{-i(\theta_k-\theta_+)}}{2(k+|\vec{k}+\vec{q}|)}+\frac{1}{2(k+|\vec{k}+\vec{q}|)}}\\ \nonumber
&&\int \frac{d^2k'}{(2\pi)^2}U(\vec{k}-\vec{k}')\left[ \tilde{\chi}_{11}(\vec{k}',\vec{q})-e^{-i(\theta_k-\theta_+)}\tilde{\chi}_{22}(\vec{k}',\vec{q})\right],\\
\lefteqn{\tilde{\chi}_{22}(\vec{k},\vec{q})=\frac{1-e^{i(\theta_k-\theta_+)}}{2(k+|\vec{k}+\vec{q}|)}+\frac{1}{2(k+|\vec{k}+\vec{q}|)} }\\ \nonumber
&&\int \frac{d^2q'}{(2\pi)^2}U(\vec{k}-\vec{k}')\left[-e^{i(\theta_k-\theta_+)}\tilde{\chi}_{11}(\vec{k}',\vec{q})+\tilde{\chi}_{22}(\vec{k}',\vec{q})\right],
\end{eqnarray}
where $\theta_k$ and $\theta_+$ are the angular coordinates of the two-dimensional momenta $\vec{k}$ and $\vec{k}+\vec{q}$ respectively. Without loss of generality, we can set $q_y=0$,
from which
it is easy to see that $\tilde{\chi}_{22}(k_x,-k_y;\vec{q})=\tilde{\chi}_{11}(k_x,k_y;\vec{q})$.
When this is substituted into the last of Eqs. \ref{insertion} one readily sees
that this insertion vanishes.  Thus our approximation for $M(\vec{q})$ includes
{\it only} the irreducible ladder diagrams.

The calculation of the $M(\vec{q})$ follows steps very analogous to those described
for the 4-leg vertex, and were outlined in Ref. \onlinecite{asym}.
The equation for the 3-leg antisymmetric vertex [Fig.~\ref{fig:3leg_diagram}] is
\begin{align}
\tilde{\Gamma}^M_{\alpha\beta}({\vec k},{\vec q})=&\, \sigma^z_{\alpha\beta}+\int\frac{d^2q'}{(2\pi)^2}U(|\vec{q'}|)\tilde{K}_{\alpha\beta\gamma\delta}(\vec{k}-\vec{q}',\vec{q}) \notag \\
& \times \tilde{\Gamma}^M_{\gamma\delta}(\vec{k}-\vec{q}',\vec{q}).
\end{align}
%\cite{no_screening},
%where
%\begin{equation}
%\tilde{K}_{\alpha\beta,\mu\nu}(\vec{p},\vec{q})\equiv %i\int\frac{dp_0}{2\pi}G^{(0)}_{\mu\alpha}(p+q)G^{(0)}_{\beta\nu}(p).
%\end{equation}
Defining
%\begin{equation}
$
\tilde{\chi}^M_{\alpha\beta}(\vec{k},\vec{q})=\tilde{K}_{\alpha\beta\gamma\delta}(\vec{k},\vec{q})\tilde{\Gamma}^M_{\gamma\delta}(\vec{k},\vec{q}),
$
%\end{equation}
one finds
\begin{align}
\tilde{\chi}^M_{\alpha\beta}(\vec{k},\vec{q})=&\, \tilde{K}_{\alpha\beta\gamma\delta}(\vec{k},\vec{q})\sigma^z_{\gamma\delta}+\tilde{K}_{\alpha\beta\gamma\delta}(\vec{k},\vec{q})\notag \\
&\times \int\frac{d^2q'}{(2\pi)^2}U(|\vec{q}'|)\tilde{\chi}^M_{\gamma\delta}(\vec{k}-\vec{q}',\vec{q}).
\end{align}
This quantity is related to the susceptibility by
\begin{equation}
M(\vec{q})=\int\frac{d^2k}{(2\pi)^2}\sigma^z_{\alpha\alpha}\tilde{\chi}^M_{\alpha\alpha}(\vec{k},\vec{q}).
\end{equation}

\subsection{Solutions at Long Wavelengths}

In what follows we focus on the long wavelength limit (small $q$), so we
drop all terms of $O(q^2)$ and higher.  Using a circular
moment expansion one finds
\begin{align}
\tilde{\chi}^{M(0)}_{\underline{\alpha\alpha}}(k,\vec{q})=&\,\tilde{K}^{(0)}_{\underline{\alpha\alpha}\beta\beta}\sigma^z_{\beta\beta}
+\tilde{K}^{(0)}_{\underline{\alpha\alpha}\beta\beta}(k,\vec{q})\frac{\beta}{k}\notag \\
&\times \int^\Lambda_{k_0} k'dk'f_0\left(\frac{k'}{k}\right) \tilde{\chi}^{M(0)}_{\beta\beta}(k',\vec{q})\label{gen_q=0_eq}.
\end{align}
Here we used the superscript $(0)$ to denote the circular component
$m=0$, and the underlined indices are not summed over.  Note that
we have introduced both an ultraviolet cutoff ($\Lambda\sim
2\pi/a$, $a$ = lattice spacing) and an infrared cutoff ($k_0\sim 2\pi/L$, $L$ =
linear size of system).

Defining $
\tilde{\chi}^{M(0)}(k,\vec{q})\equiv\sigma^z_{\beta\beta}\tilde{\chi}^{M(0)}_{\beta\beta}(k,\vec{q}),
$ in the limit $q\rightarrow0$ the solution to Eq.~(\ref{gen_q=0_eq})
may be written in the form
$\tilde{\chi}^{M(0)}(k,0)=\frac{1}{v_Fk}F\left(\frac{k}{\Lambda}\right),$
where $F$ obeys the integral equation
\begin{equation}
F\left(\frac{k}{\Lambda}\right)=1+\frac{\beta}{2k}\int^\Lambda_{k_0}dk'f_0\left(\frac{k'}{k}\right)F\left(\frac{k'}{\Lambda}\right).
\label{Fwithq0}
\end{equation}
Note that $F$ depends on the ratio $k/\Lambda$, a
reflection of the fact that the original Hamiltonian has no intrinsic
length scale, so (in the limit $k_0 \rightarrow 0$) $k$ can enter only
in this ratio.  For $k/\Lambda\ll 1$, one easily confirms that
Eq. (\ref{Fwithq0}) is solved by a power law $F\left(\frac{k}{\Lambda}\right) \sim (\Lambda/k)^s$,
with $s$ going from
real to complex above some critical $\beta$.  This is precisely the
behavior we identified in the 4-leg vertex; unlike what one finds in the density-density
response case, the power law is not canceled upon forming the 3-leg vertex
from the 4-leg vertex.

\begin{figure}
\begin{center}
\subfigure[]{\label{fig:Fkq0b03}\includegraphics[scale=0.65]{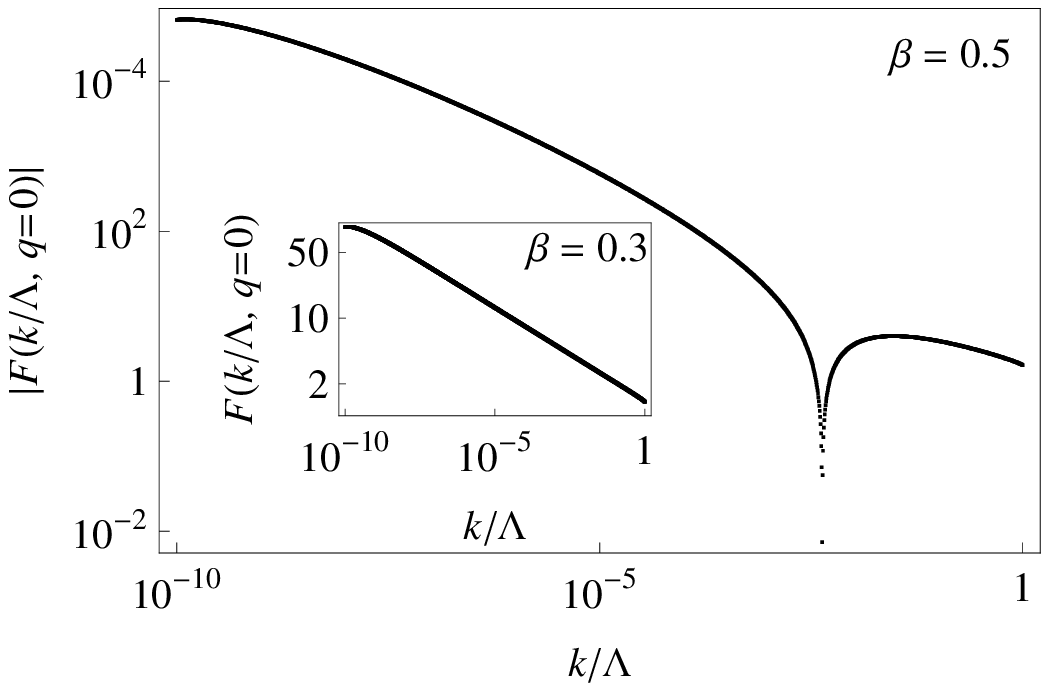}}
\subfigure[]{\label{fig:Mq0}\includegraphics[scale=0.65]{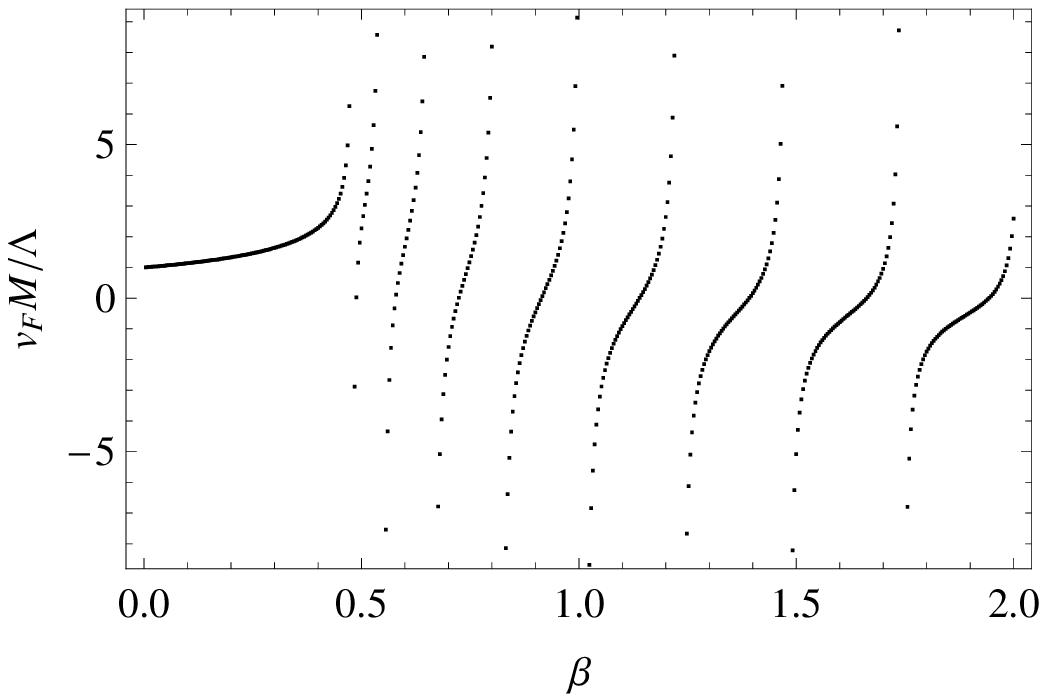}}
\end{center}
\caption{\label{q0results}Solutions of Eq.~(\ref{Fwithq0}) with $k_0/\Lambda=10^{-10}$. (a) $\beta=0.5$. Because the plotted is $|F|$, the oscillations appear as cusps. Note the amplitude of the oscillation scales roughly as $1/\sqrt{k}$. {\it Inset:} $F$ for $\beta=0.3$. It is clearly a power law except for $k$ close to $k_0$. (b) The antisymmetric response $M$ as a function of the interaction strength $\beta$.
}
\end{figure}

Eq.~(\ref{Fwithq0}) may be readily solved numerically. For small
$\beta$, the solution is indeed a power law, provided $k\gg k_0$ [see
Fig.~\ref{fig:Fkq0b03} inset].  For large enough $\beta$, the
solution is consistent with a power law of complex exponent, such that
$F$ becomes oscillatory with a power law envelope
[Fig.~\ref{fig:Fkq0b03}].
Interestingly,
$ M(\vec{q}\rightarrow
0)=\int\frac{d^2k}{(2\pi)^2}\tilde{\chi}(k,\vec{q}\rightarrow 0) $ also has
a series of divergences [Fig.~\ref{fig:Mq0}].
Formally, one may understand the occurrence of these
poles by thinking
of the
solution in terms of the inverse of $1-\beta\hat{L}$, where $\hat{L}$
is the integral operator on the right hand side of
Eq. (\ref{Fwithq0}). Divergences then occur as $\frac{1}{\beta}$
crosses successive eigenvalues of $\hat{L}$.
The presence of such poles suggests a phase transition into
a state with a spontaneously generated $M(q \rightarrow 0)$
becoming a Dirac mass, i.e., chiral symmetry breaking.
However, the positions and weights of these poles are
sensitive to $k_0$, the infrared cutoff due to the finite
system size, and merge together in the $L \rightarrow \infty$
limit to introduce
a branch cut in $F$ as a function of $\beta$.  We discuss
the significance of this below.

\begin{figure}[t]
\begin{center}
\includegraphics[scale=0.75]{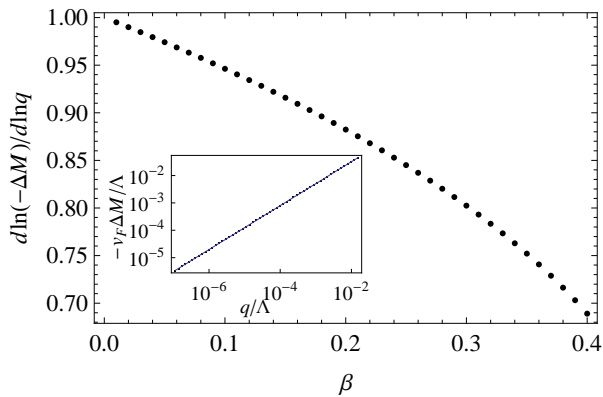}
\end{center}
\caption{\label{fig:dM}The exponent in $\Delta M(q)$ as a funciton of $\beta$. {\it Inset:} $\Delta M(q)$ for $\beta=0.3$.
}
\end{figure}

For small but nonzero $q$, it is interesting to compute the correction
$\Delta M(q)=M(q)-M(0)$.  The equation for the corresponding $\Delta
F$ has a form very similar to Eq.~(\ref{Fwithq0}), with only the
``$1$'' replaced by an inhomogeneous term, which is proportional to
$q^2/k^2$ for $k \gg q$.  The $\Delta M(q)$ resulting from this then
vanishes with an exponent that varies with $\beta$.  The inset of Fig.~\ref{fig:dM}
illustrates a typical result for $\beta$ not too large; the
exponent as a function of $\beta$ is illustrated in
the main panel of Fig.~\ref{fig:dM}.
One physical consequence of this is that
the difference in charge between
sublattices for an impurity placed asymmetrically with respect
to the sublattices will fall off with a $\beta$-dependent power law at
large distances, behavior which
may be observable with a local scanning probe.
%We note that $\Delta M(q)$
%has singularities at the same values of $\beta$ as $M(0)$, as should
%be expected from the form of Eq.~(\ref{Fwithq0}).

The result illustrated in
Figs. \ref{q0results} and \ref{fig:dM}
have a number of interesting consequences for
interacting electrons in undoped graphene.
For $\beta<\beta_c$, we see that there are indeed
power law correlations at long distances
in quantities that are in principle measurable,
with an exponent varying continuously
with $\beta$.  This means that
the weak-coupling many-body groundstate  possesses a basic
property of a critical phase.
For $\beta>\beta_c$ the exponent becomes complex, as in the
noninteracting Coulomb implosion problem. In the interacting many-body
case, the susceptibility $M(q)$ of
Eq. (\ref{M(q)}) diverges for $k_0 >
0$. This strongly suggests a quantum phase transition
to broken symmetry state with
staggered charge
order \cite{Khveshchenko,Gorbar2002}.

The evolution
of this system, from a state with power law correlations
to one in which a transition occurs when this power
reaches some limiting value, is highly reminiscent
of the phenomenology of systems which undergo a KT transition \cite{nelson_2002}.
In such systems the transition indicates the appearance
of a correlation length, which equivalently indicates
that a gap develops in the excitation spectrum.  This behavior
in general should be signaled by a divergence in an appropriate response function.
However, the presence of {\it many} such divergences
as a function of $\beta$ suggests there are different
ways to break the symmetry.
For a system with finite system size, we expect this
chiral symmetry-breaking to occur as $\beta$ increases
from small values in a way that is consistent with the first
such pole.  As we show below,
the separation between neighboring poles
vanishes only
logarithmically as
$k_0 \rightarrow 0$, so that finite size may
in fact be important for realistic system sizes.

Nevertheless, the merging of these poles suggests
that something else must happen
in the thermodynamic limit \cite{ref_credit}.
The merging of these poles as
$k_0 \rightarrow 0$ results in a
a continuous function with a
branch point at $\beta_c$.  We interpret this latter non-analytic behavior
as the signal of a phase transition.
Since it is a result of the merging poles, a natural interpretation is
that the instability is into a state involving fluctuations among
different realizations of a chiral order parameter which, if quiescent, would
produce a gapped exciton phase \cite{Khveshchenko,Gorbar2002}.
We speculate that with further increase in $\beta$, one of these orderings
could be favored over the others, resulting in a true condensed phase.
This would be consistent with results of quantum Monte Carlo calculations \cite{drut123}.

\subsection{An Analytical Model}

A fuller understanding of Eq.~(\ref{Fwithq0}) may be arrived at
with a model
kernel of the form
\begin{equation}
\tilde{f}_0(x)=\theta(1-x)+\frac{1}{x}\theta(x-1).
\end{equation}
This has the same behavior as the real kernel at large and small $x$,
and is simple enough to allow analytic solutions.
We have verified numerically that the results for $F$ and $M$ are
qualitatively very similar to those obtained with the correct $f_0$.

\subsubsection{\label{sec:into}Solution by Wiener-Hopf Method}
%Section on the Wiener-Hopf solution of the integral equation with a
%simplified kernel. Put it wherever you want, Herb and Jianhui.

With this model kernel, assuming that the integral converges as
$q\to 0$, which will be checked after the solution is found, we obtain,
in the $k_0 \to 0$ limit,
the integral equation
\begin{equation} F\bigg(\frac{k}{\Lambda}\bigg)-\frac{\beta}{2k}\int\limits_{0}^{\Lambda} dk^{\prime} {\tilde{f}}_0\bigg(\frac{k^{\prime}}{k}\bigg) F\bigg(\frac{k^\prime}{\Lambda}\bigg)=1.
\label{wh1}\end{equation}
Let us change to more convenient variables via
\begin{equation}
\frac{k}{\Lambda}=e^{-t},\ \ \ \ \frac{k^{\prime}}{\Lambda}=e^{-t^{\prime}}
\label{wh2}\end{equation}
and rename the function for which we are solving, as well as the kernel,
\begin{equation}
F\bigg(\frac{k}{\Lambda}\bigg)=g(t),
\ \ \ \ {\tilde{f}}_0\bigg(\frac{k^{\prime}}{k}\bigg)=K(t-t^{\prime}),
\label{wh3}\end{equation}
to obtain the rewritten integral equation
\begin{equation}
g(t)-\frac{\beta}{2} \int\limits_{0}^{\infty} dt^{\prime} e^{t-t^{\prime}}K(t-t^{\prime})g(t^{\prime})=1.
\label{wh4}\end{equation}

Note that physically meaningful values of $t$ are nonnegative.
An important point is that if the limits of integration over $t^{\prime}$
had been $(-\infty,\infty)$ one could have solved the equation
trivially by Fourier transformation. Since it is over the half-line
one has to use the more sophisticated Wiener-Hopf method \cite{hazewinkel_2001}.

One first extends the definition of $g$ so that it has the entire real
line for its domain, defining
\begin{equation}
g(t)=g_+(t)+g_-(t),
\label{wh5}\end{equation}
where $g_+(t)$ is nonzero only when $t\ge 0$ and
$g_-(t)$ is nonzero only for $t\le0$. As part of the
solution one obtains both $g_+$ and $g_-$. Defining
$R(t)=e^{t}K(t)$ we can now extend the range of
integration over $(-\infty,\infty)$ as long as one integrates only
$g_+$:

\begin{equation}
g_+(t)-\frac{\beta}{2}\int\limits_{-\infty}^{\infty} dt^{\prime} R(t-t^{\prime})g_+(t^{\prime})=1-g_-(t)=r_+(t)+r_-(t)
\label{wh6}\end{equation}
where $r_+(t)=\Theta(t)$ and $r_-(t)=\left[1-g_-(t)\right]\Theta(-t)$.

One can now solve the equation by Fourier transformation.  The crucial
point is that since $g_+$ is nonzero only for nonnegative values, if
it vanishes as $t\to\infty$, its Fourier transform has poles only
in the lower half-plane of complex $\omega$, while $g_-$ has poles
only in the upper half-plane. This gives us the extra information
needed to solve for both $g_+$ and $g_-$. In general, there is no need
for $g_+$ to vanish as $t\to\infty$, which would correspond to
the original function $F$ vanishing as $k\to0$. In fact, one expects
$F$ to diverge with a power law as $k\to0$. To incorporate this
expectation, we define $g_+(t)=e^{st}h_+(t)$, with
$h_+$ vanishing as $t\to\infty$. The new function $h_+$ satisfies an integral equation with a modified kernel $R_s(t)=e^{-st}R(t)$

\begin{equation}
h_+(t)-\frac{\beta}{2} \int\limits_{-\infty}^{\infty}dt^{\prime}R_s(t-t^{\prime})h_+(t^{\prime})=e^{-st}(r_+(t)+r_-(t)).
\label{wh7}\end{equation}

We now take the Fourier transform of both sides, using the explicit
form of $\tilde{f}_0$, which corresponds to
\begin{equation}
R_s(t)=e^{(1-s)t}\Theta(-t)+e^{-st}\Theta(t).
\label{wh8}\end{equation}

We will abuse notation slightly by using the same name for the
function and its Fourier transform, the argument and context serving
to distinguish them. Thus, $R_s(t)$ has the Fourier transform

\begin{equation}
R_s(\omega)=\int\limits_{-\infty}^{\infty} dt e^{i\omega t} R_s(t)=\frac{1}{i\omega+1-s}+\frac{1}{-i\omega+s}.
\label{wh9}\end{equation}
The existence of the Fourier transform implies $0\le s\le 1$, but
does not choose $s$ uniquely. In general, the choice of $s$ determines
the class of functions which are allowed as solutions, as we will see
explicitly below.  Now the equation becomes
\begin{eqnarray}
&\bigg(\frac{(i\omega)^2+i\omega(1-2s)+\beta/2-s(1-s)}{(i\omega-s)(i\omega+1-s)}\bigg)h_+(\omega)\nonumber\\
&=-\frac{1}{i\omega-s}+r_-(\omega+is).
\label{IE-FT}
\end{eqnarray}

To proceed further we separate the prefactor of $h_+$ on the left hand
side [call it $P(\omega)$] into a {\it product}
$P(\omega)=P_+(\omega)P_-(\omega)$, where, by construction,
$P_+(\omega)$ has zeroes and poles only in the lower half-plane and
$P_-(\omega)$ has zeroes and poles only in the upper half-plane.
We denote the roots of the numerator of $P$ (as a function of $i\omega$) as
\begin{equation}
x_{\pm}=s-\frac{1}{2}\pm \frac{1}{2}\sqrt{1-2\beta}.
\label{wh11}\end{equation}
One possible choice of $s$ is to make $x_+\ge0,\ \ x_-\le0$. Let us
analyze this case first.  This allows us to determine $P_{\pm}$ uniquely.

\begin{equation}
P_+(\omega)=\frac{i\omega-x_+}{i\omega-s},\ \ \ \ P_-(\omega)=\frac{i\omega-x_-}{i\omega+1-s}.
\label{wh12}\end{equation}

Now divide through Eq. (\ref{IE-FT}) by $P_-(\omega)$ to obtain

\begin{equation}
P_+(\omega)h_+(\omega)=-\frac{i\omega+1-s}{(i\omega-x_-)(i\omega-s)}+
\frac{r_-(\omega+is)(i\omega+1-s)}{i\omega-x_-}.
\label{wh13}\end{equation}
The first term on the right hand side has poles in both half-planes,
and we separate them by partial fractions

\begin{eqnarray}
P_+(\omega)h_+(\omega)=&-\frac{1}{(s-x_-)(i\omega-s)}\nonumber\\
+\frac{1+x_--s}{(s-x_-)(i\omega-x_-)}&+\frac{r_-(\omega+is)(i\omega+1-s)}{i\omega-x_-}.
\label{wh14}\end{eqnarray}
Since the product $P_+h_+$ is guaranteed by construction to have poles
only in the lower half-plane the terms with the poles in the upper
half-plane on the right hand side must separately vanish, and we
obtain

\begin{equation}
h_+(\omega)=-\frac{1}{(s-x_-)(i\omega-x_+)}
\label{wh15}\end{equation}

Going back to the fictitious ``time'' variable $t$ and using  $g_+(t)=e^{st}h_+(t)$, we obtain

\begin{equation}
 g_+(t)=\frac{\Theta(t)e^{(s-x_+)t}}{(s-x_-)}
\label{wh16}\end{equation}
Note that with the definitions of $x_{\pm}$, $s$ drops out of this expression.
Translating back to the original variables, we see that $F\propto
\big(\frac{k}{\Lambda}\big)^{-\frac{1}{2}+\frac{1}{2}\sqrt{1-2\beta}}$.
%In hindsight, one could have chosen
%$s$ to satisfy $s(1-s)=\beta$ to simplify the intermediate steps.
It
is easily verified that, as assumed in Eq. (\ref{wh1}), the integral
converges as $k\to 0$.  Thus, the Wiener-Hopf method demonstrates the power
law behavior of the solution to the integral equation.

Another choice of $s$ would be to make both $x_{\pm}>0$,
yielding a solution which diverges more slowly as $k \to 0$.
The general solution is a linear combination of both these
solutions, and interestingly we see that the equation in
its present form does not uniquely specify a particular combination.
This ambiguity is lifted by introducing a lower cutoff $k_0$ in momentum,
corresponding to a finite system size.  We show below using an alternate
method how this leads to a unique solution.

\subsubsection{Solution of Equivalent Differential Equation}

Eq.~(\ref{Fwithq0}) can also be solved with the model kernel
by converting it into a differential equation.

Differentiating Eq.~(\ref{Fwithq0}) we get
\begin{align}
F'(k)=&\frac{\beta}{2}[-\frac{1}{k^2}\int_{k_0}^1dk'\tilde{f}_0\left(\frac{k'}{k}\right)F(k') \notag \\
&+\frac{1}{k}\int_k^1dk'\frac{1}{k'}F(k')]\label{eq-F-prime}.
\end{align}
Differentiating this equation, we get
\begin{align}
F''(k)=&\frac{\beta}{2}[\frac{2}{k^3}\int_{k_0}^1dk'\tilde{f}_0\left(\frac{k'}{k}\right)F(k')-\frac{2}{k^2}\int_k^1dk'\frac{1}{k'}F(k')\notag \\
&-\frac{1}{k^2}F(k)],
\end{align}
but from Eq.~(\ref{eq-F-prime}), the first 2 terms in the square brackets are simply $-\frac{2}{k} \frac{2F'(k)}{\beta}$. Therefore the differential equation corresponding to Eq.~(\ref{Fwithq0}) is
\begin{equation}
F''(k)+\frac{2}{k}F'(k)+\frac{\beta}{2k^2}F(k)=0,
\end{equation}
which has general solutions of the form
\begin{equation}
F(\tilde{k})=A_+ \tilde{k}^{\lambda_+}+A_- \tilde{k}^{\lambda_-},
\label{model_F}
\end{equation}
with $\tilde{k}=k/\Lambda$,
$\lambda_{\pm}=\frac{-1\pm\gamma}{2}$, and $\gamma=\sqrt{1-2\beta}$.
The coefficients $A_{\pm}$ are determined by substituting Eq. \ref{model_F}
back into the integral equation.  This results in power law behavior
for $k \gg k_0$, with exponent $\lambda_+$, which goes from real to
complex when $\beta$ exceeds 1/2.  Moreover, $M(q \rightarrow 0)$ may
be evaluated, yielding
\begin{equation}
M(0)=\frac{\Lambda}{v_F}\frac{2-2\tilde{k}_0^\gamma}{1+\gamma-\beta+\tilde{k}_0^\gamma (-1+\gamma+\beta)}.
\end{equation}
This
has poles for $\beta>1/2$ when
\begin{equation}
\sqrt{2\beta-1}\ln{\tilde{k}_0}=2\arctan{\frac{\sqrt{2\beta-1}}{1-\beta}}+2\pi n,
\end{equation}
with integer $n$ and $0<\arctan{(x)}<\pi$.  Note that the
distance between poles vanishes logarithmically as $\tilde{k}_0 \rightarrow
0$, as discussed above.
%\textbf
{Furthermore, for $\beta>1/2$, $\tilde{k}_0^{\gamma}$
becomes ill-defined unless an infinitesimal imaginary part is introduced
in $\beta$, so that $\beta=1/2$ becomes a branch point for $M(0)$.  We interpret
this as the signal of a phase transition in the thermodynamic limit, since
$M(0)$ need not be real and positive beyond this point.}

\section{\label{sec-den}density response}

In this final section we return to another measurable quantity,
the density response function.  To be concrete we will use our
result to compute the induced charge around a Coulomb impurity,
which generates the potential
$Ze/\epsilon r$, where $e\equiv |e|$.
Our procedure is to
first solve
Eq.~(\ref{eq-tild-Gam}) numerically,
from which we compute the (static) irreducible polarizability
\begin{equation}
\Pi(\vec{q})=i\int\frac{d^2k}{(2\pi)^2}\tilde{K}_{\beta_2\beta_1\alpha\alpha}\tilde{\Gamma}_{\beta_1\beta_2}(\vec{k},\vec{q}).
\end{equation}
The density response function is then computed by an RPA sum \cite{fetter},
except that instead of using the non-interacting polarizability we
use our irreducible polarizability, which includes excitonic corrections
via the ladder diagrams.  The result of this takes the form
\begin{equation}
D(q)=\frac{-\Pi(q)}{1+\Pi(q)U_C(q)},
\end{equation}
where $U_C(q)=2\pi\beta v_F/q$ is the Coulomb interaction.
Finally, the Fourier transform of the induced electron density is given by
\begin{equation}
\delta n(q)=D(q) \phi_{\text{ext}}(q).
\end{equation}
In the Coulomb impurity case, the external potential $\phi_{\text{ext}}(q)=-ZU_C(q)$.

To find numerical solutions to Eq.~(\ref{eq-tild-Gam}),
we discretize the allowed values of momentum $\vec{k}$ and $\vec{q}$,
and replace integrations by sums over the grid of allowed momenta.
In doing this, some
subtleties arise. Since we cannot retain an infinite
number of momentum points, we must confine the sums to
a finite region, most conveniently taken to be square.
If we use the MDF spectrum and wavefunctions
(needed to construct $\tilde{K}$) in Eq. \ref{eq-tild-Gam}
in this
``Brillouin zone'', the former will be periodic
but not the latter.  The discontinuity in wavefunctions
leads to spurious oscillations in the final result.
In principle this can be overcome by simulating the
system on a honeycomb lattice and using the full
tight-binding spectrum and wavefunctions for graphene.
However, this is numerically costly and unnecessary, because
the low energy physics is almost entirely determined
by the wavefunctions and spectra near the Dirac cones. Moveover,
since Coulomb interactions are relatively weak at
short wavelengths,
one can neglect the intervalley scattering, so that it
should be sufficient to consider only one Dirac cone, whereas
a simulation of a honeycomb lattice would force us to
include two due to fermion doubling \cite{Castro_Neto_RMP}.

As a compromise
we consider models that have
simpler bandstructures than graphene but still
have a Dirac cone. One such model
arises in the theory of the surface of a topological insulator
\cite{Qi2004,Qi2008},
and has the form ($\hbar v_F=1$)
\begin{align}
\label{eq-model}
H=&\sum_{n}\left[\vec{c}_n^\dagger\frac{\sigma_z-i\sigma_x}2\vec{c}_{n+\hat{x}}+
\vec{c}_n^\dagger\frac{\sigma_z-i\sigma_y}2\vec{c}_{n+\hat{y}}+h.c.\right]\\ \nonumber
& +m\sum_n \vec{c}_n^\dagger \sigma_z\vec{c}_n \\
=&\sum_{\vec{k}}\vec{c}^{\dagger}_{\vec{k}}h(\vec{k})\vec{c}_{\vec{k}},
\end{align}
where
\begin{equation}
\label{eq-hk}
h(\vec{k})=(\sin k_x)\sigma_x+(\sin k_y)\sigma_y+\left(m+\cos k_x+\cos k_y\right)\sigma_z.
\end{equation}
This is a model defined on a square lattice,
with each site supporting a two-component vector
of localized orbitals combined into annihilation operators $\vec{c}_n$, and
with the sum over $n$
running through all the lattice sites.
The crystal momentum $\vec{k}$ is measured in units
of $1/a$, with $a$ the lattice constant.
For $m=2$, there is a single Dirac point at the center of the BZ,
and the spinor structure in the vicinity of this point
is the same as near the Dirac points in graphene.
Thus we expect this model to reproduce the low-energy behavior of graphene.
We adopt this model for our numerical solution of
Eq.~(\ref{eq-tild-Gam}).

A second subtlety arises in the doped case.
For example, when the chemical potential $\mu>0$,
the quantity $\tilde{K}$ in Eq.~(\ref{eq-tild-Gam}) takes the form
\begin{align}
& \tilde{K}^{\mu>0}_{\alpha_1\beta_1,\alpha_2\beta_2}(\vec{k},\vec{q})=\label{eq-K-mu-gt-0}\\ \nonumber
& \frac{\left[\theta(\varepsilon_{\vec{k}}-\mu)- \theta(\varepsilon_{\vec{k}+\vec{q}}-\mu)\right]g^+_{\beta_1\beta_2}(\vec{k})g^+_{\alpha_2\alpha_1}(\vec{k}+\vec{q})}{\varepsilon_{\vec{k}}-\varepsilon_{\vec{k}+\vec{q}}}\\ \nonumber
& +\frac{\theta(\varepsilon_{\vec{k}}-\mu)g^+_{\beta_1\beta_2}(\vec{k})g^-_{\alpha_2\alpha_1}(\vec{k}+\vec{q})}{\varepsilon_{\vec{k}}+\varepsilon_{\vec{k}+\vec{q}}}\\ \nonumber
& +\frac{\theta(\varepsilon_{\vec{k}+\vec{q}}-\mu)g^-_{\beta_1\beta_2}(\vec{k})g^+_{\alpha_2\alpha_1}(\vec{k}+\vec{q})}{\varepsilon_{\vec{k}}+\varepsilon_{\vec{k}+\vec{q}}},
\end{align}
where $\varepsilon_{\vec{k}}$ is the positive eigenvalue of $h(\vec{k})$ and $g^{\pm}_{\alpha\beta}(\vec{k})=(\eta^{\pm}_{\vec{k}})_{\alpha}(\eta^{\pm}_{\vec{k}})_{\beta}^*$, with $\eta^{\pm}_{\vec{k}}$ being the eigenvectors of $h(\vec{k})$ corresponding to $\pm \varepsilon_{\vec{k}}$ respectively.

Because of the step functions, a naive numerical
integration by a discrete summation works
poorly, because the function being integrated
is not smooth on the scale of the grid. This problem
may be overcome using the
Triangular Linear Analytic (TLA) method\cite{first-TLA,second-TLA}.
We divide the square Brillouin zone into small squares,
and each small square is further subdivided
into two right trangles along
one of the diagonals.  Weights for the integrand
can then be assigned
at the corners of the triangles employing the parameterization
formulas in Ref.~\onlinecite{second-TLA} \cite{triangles_fixed}.
Using this weighting scheme to approximate the integrals
gives far better results than a naive lattice sum.

Finally, we note that $U(q)$ in Eq.~(\ref{eq-tild-Gam}) is
better represented by
the RPA screened Coulomb interaction than the bare
Coulomb interaction, i.e.,
\begin{equation}
U(q)=\frac{U_C(q)}{1+\Pi^{\text{RPA}}(q)U_C(q)}.
\end{equation}
The irreducible RPA polarizability $\Pi^{\text{RPA}}(q)$
has been calculated by a number of authors
(see, e.g. Refs.~\onlinecite{Hwang-polarizability} and \onlinecite{wunsch_2006}).
For the case of undoped graphene there is no qualitative difference
between using screened or unscreened Coulomb interactions
in the ladder rungs, because the functional forms of
$U(q)$ and $U_C(q)$ are the same, and only the effective
value of $\beta$ is renormalized.  When $\mu \ne 0$, however,
the Fermi surface introduces a length scale into the problem,
allowing genuine screening of the Coulomb interaction at long
distances.  This can be modeled by using a contact interaction
on the ladder rungs \cite{brey_1989,wunsch_2006}
although we find this introduces problems at large wavevectors,
as we describe below.

\begin{figure}
\begin{center}
\includegraphics[width=0.75\columnwidth]{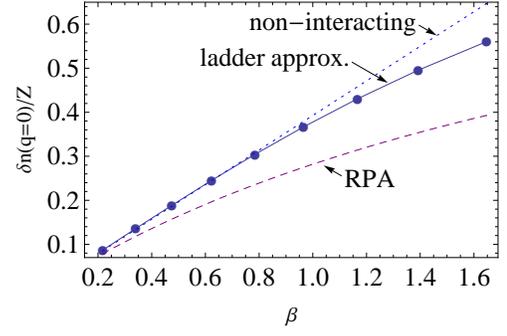}
\end{center}
\caption{\label{fig-tot-charge-undoped} (Color online) Total induced electron number density divided by $Z$ in the undoped case.}
\end{figure}

Fig.~\ref{fig-tot-charge-undoped} shows the total induced electron
number density in the undoped case, together with the RPA result
\begin{equation}
\frac{\delta n^{\text{RPA}}(q=0)}{Z}=\frac{\Pi^{\text{RPA}}(q)U_C(q)}{1+\Pi^{\text{RPA}}(q)U_C(q)}
 =\frac{\frac{\pi \beta}{8}}{1+\frac{\pi \beta}{8}}
\end{equation}
and the non-interacting result
\begin{equation}
\frac{\delta n^{\text{non}}(q=0)}{Z}=\Pi^{\text{RPA}}(q)U_C(q)=\frac{\pi \beta}{8}
\end{equation}
for comparison. It is clear that the non-interacting
result exceeds 1 for $\beta>8/\pi$, while the RPA result
approaches 1 as $\beta \rightarrow \infty$.
Results for the total induced charge using the RPA and the
ladder approximation are not qualitatively different.
The ladder approximation result is larger than the RPA result, and the difference increases with $\beta$.

\begin{figure}
\begin{center}
\includegraphics[width=0.9\columnwidth]{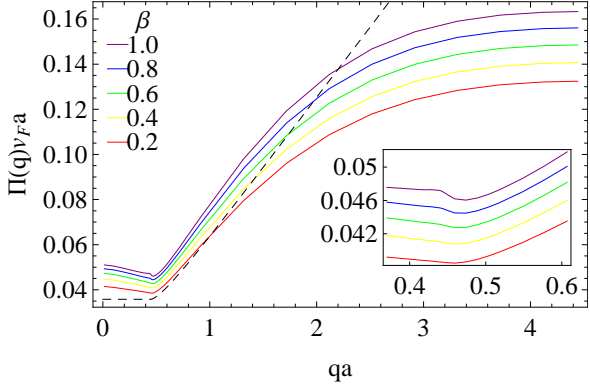}
\end{center}
\caption{\label{fig-Pi-TI-doped} (Color online) The continuous color curves are the irreducible polarizability $\Pi(q)$ for the model Hamiltonian Eq.~(\ref{eq-model}) in the doped case, with RPA screened Coulomb interaction as the rungs of the ladders, for $\beta=0.2,0.4,\dots,1.0$, with higher curves corresponding to higher $\beta$. The dashed black curve is the RPA result for MDF for comparison. The inset is a blow-up showing the developing structure just below $q=2k_F$.
For all the numerical results shown in this figure $\mu=0.225\hbar v_F/a$.}
\end{figure}

For doped graphene, any charged impurity will induce
an equal and opposite screening charge, both in the
RPA and when exciton corrections are included.
However, we find in the latter case a strong
quantitative difference between the two in
the shape of the screening cloud.
This is due to the effect of the exciton corrections
on the irreducible
polarizability in the doped case, illustrated in
Fig.~\ref{fig-Pi-TI-doped}. For small $\beta$,
$\Pi$ is close to the RPA result for MDF as expected,
except that the for $q<2k_F$ there is a negative slope,
and for $q$ close to the Brillouin zone boundary it is
significantly below the RPA result for MDF, simply
because of the presence of the zone boundary. For
larger $\beta$, the curve is higher and the deviation
from the RPA result for MDF is larger, and for
the largest $\beta$, an additional ``hump'' structure
develops just below $2k_F$, as illustrated in the
inset of Fig.~\ref{fig-Pi-TI-doped}.  We will consider
this extra structure in more detail below.

\begin{figure}
\begin{center}
\includegraphics[width=0.9\columnwidth]{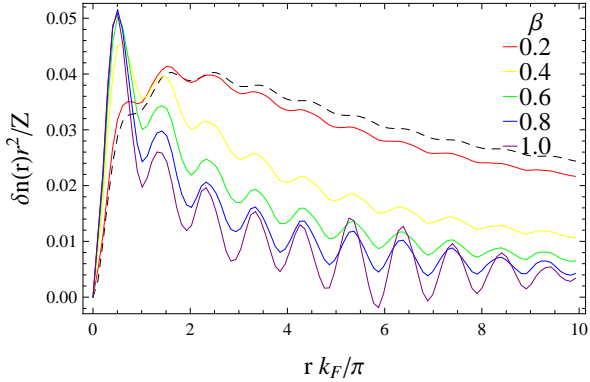}
\end{center}
\caption{\label{fig-dn-TI-doped} (Color online) Induced electron number density (times $r^2/Z$) as a function of distance $r$ from the impurity in the doped case. On the far right side, from top to bottom, the curves correspond to $\beta=0.2,0.4,\dots,1.0$. For comparison, the dashed line is the RPA result with $\beta=0.2$.}
\end{figure}

The cusp in the density response function at $q=2k_F$ is
well-known to induce Friedel oscillations \cite{friedel}.
The change in shape of the irreducible polarizability
near $2k_F$ essentially deepens this cusp, leading to
an enhancement of these oscillations.
Fig.~\ref{fig-dn-TI-doped} shows the induced charge
distribution, illustrating this enhancement.
We also note that with the exciton corrections included,
the induced charge density falls off somewhat faster than the RPA result.
%For larger $\beta$ the curves are more closer to each other.

We can also carry out the calculation with contact interactions
as rungs of the ladders, i.e. $U(q)=u_0$ in Eq.~(\ref{eq-tild-Gam}),
which then becomes
\begin{eqnarray}
\label{eq-tild-Gam-contact}
\lefteqn{\tilde{\Gamma}_{\alpha_2\beta_2}({\vec q})=\delta_{\alpha_2\beta_2}+}\\ \nonumber
&&u_0\left[ \int\frac{d^2q'}{(2\pi)^2}
\tilde{K}_{\alpha_2\beta_2\gamma_1\gamma_2}(\vec{q}',\vec{q})\right] \tilde{\Gamma}_{\gamma_1\gamma_2}(\vec{q}).
\end{eqnarray}
This is a considerable simplification relative to the equation
we had to solve for the RPA-screened Coulomb interaction;
we now have a simple set of linear equations for
$\tilde{\Gamma}_{\alpha_2\beta_2}({\vec q})$.
The resulting irreducible polarizability is
\begin{equation}
\label{eq-cntct-Pi}
\Pi(q)=\frac{\frac{1}{2\pi}\Pi^{\text{RPA}}(q)}{1-\frac{u_0}{4\pi}\Pi^{\text{RPA}}(q)}.
\end{equation}

\begin{figure}
\begin{center}
\includegraphics[width=0.9\columnwidth]{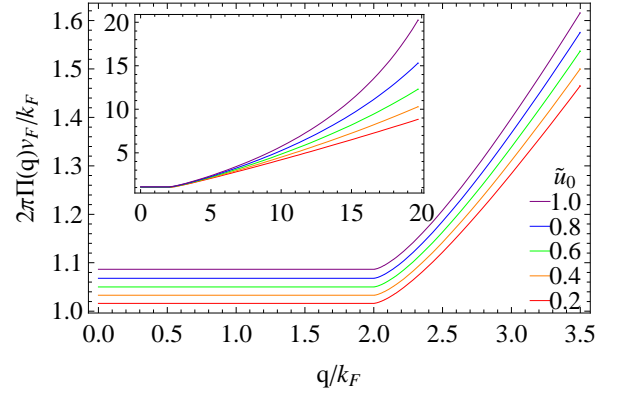}
\end{center}
\caption{\label{fig-Pi-Contact} (Color online) Irreducible polarizability in the ladder approximation with contact interaction as the rungs of the ladders, for MDF in the doped case [Eq.~(\ref{eq-cntct-Pi})]. $\tilde{u}_0 \equiv u_0 k_F/v_F$. The inset is the same plot with a larger range of $q$. }
\end{figure}

This is plotted in Fig.~\ref{fig-Pi-Contact} for the doped case.
We see that the simpler contact interaction does not enhance the
cusp at $q=2k_F$, suggesting that the correct long-distance
form of the screened Coulomb potential is an important
ingredient in obtaining this behavior.
Note also that because of the minus sign in the denominator of
Eq.~(\ref{eq-cntct-Pi}) and the monotonic increase
of $\Pi^{\text{RPA}}(q)$ at large $q$, there is a pole
at some sufficiently large $q$ for any positive $u_0$, suggesting
an instability in this model which is absent for the more
realistic $U(q)$.

\begin{figure}
\begin{center}
\subfigure[]{\label{fig-Pi-Contact-TI}\includegraphics[width=0.9\columnwidth]{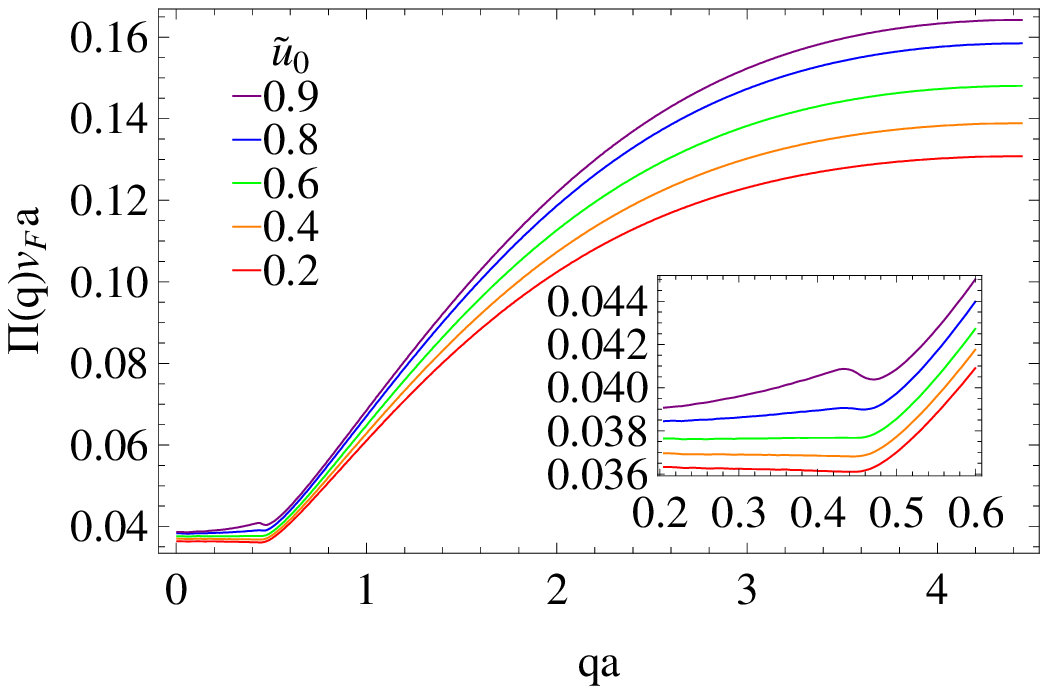}}
\subfigure[]{\label{fig-Pi-Contact-TI-u0-1}\includegraphics[width=0.9\columnwidth]{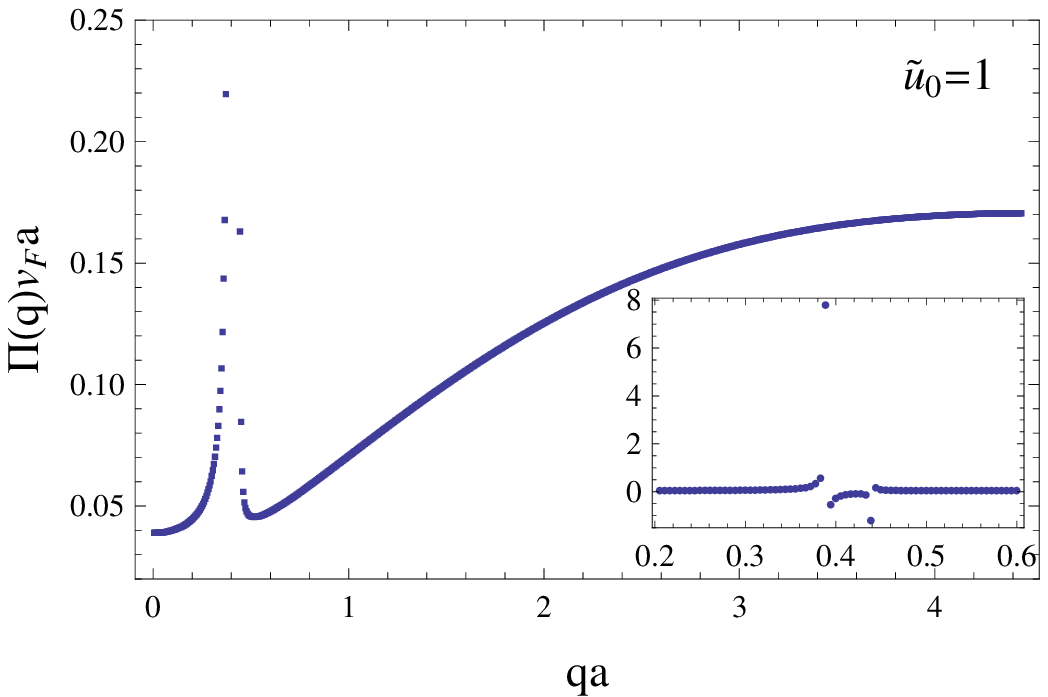}}
\end{center}
\caption{ (Color online) Numerical result for the irreducible polarizability in the ladder approximation with contact interaction as the rungs of the ladders, for the model Eq.~(\ref{eq-model}) in the doped case. The insets are the blow-up around $q=2k_F$. (a)$\tilde{u}_0=0.2,0.4,0.6,0.8,0.9$; (b)$\tilde{u}_0=1.0$. }
\end{figure}

We can also obtain
results for rungs with contact interactions numerically for
our topological insulator surface model,
Eq.~(\ref{eq-model}), illustrated in Fig.~\ref{fig-Pi-Contact-TI}.
Comparison between this and the nearly analytic results
for MDF's allow us to assess which features may be
introduced by going from the latter to the former.
As we can see, the curves are very similar in overall scale to
those when the interaction is the RPA screened
Coulomb interaction (Fig.~\ref{fig-Pi-TI-doped}).
Note however that the deepening of the $2k_F$ cusp
is {\it absent} in both the numerical result and
the analytical one, suggesting that our numerical results
are reasonably accurate at small $q$.
However, we note that for the largest values of $\beta$,
extra structure near $2k_F$ develops that appears analogous
to what we found in the Coulomb case.
This structure appears {\it only} in
the result for the model Eq.~(\ref{eq-model}),
not in the analytical result for MDF's. It is thus reasonable
to assume that the analogous structure in the RPA screened
Coulomb interaction case is peculiar to Eq.~(\ref{eq-model}) as well.
It is interesting to speculate then that one may be able
to distinguish the Dirac cone in the graphene system from
that of at the surface of a topological insulator through such structure at
very large $\beta$.

Finally, in Fig.~\ref{fig-Pi-Contact-TI-u0-1}
we illustrate that
divergent behavior emerges when $u_0$ is sufficiently
large, which evolves into a double pole from the ``hump'' structure
below $2k_F$.  This
suggests the system becomes
unstable for contact interactions of sufficiently large magnitude, a behavior
that occurs as we observed above for {\it any} $u_0$ when
MDF's are subject to contact interactions.
(Note, however, in this case the instability sets in for
$q<2k_F$, whereas for contact interactions the instability occurs
at much larger $q$ when the interaction is weak.)
While this behavior is absent in the screened Coulomb interaction
case, it is possible that at very small distances where
the atomic orbital physics becomes relevant such a contact
model becomes appropriate.  Since this instability appears in
the density-density response function this naively suggests
that there is a phase transition into a charge-density wave.
However, other transitions -- spin or valley density waves,
for example -- may preempt this transition.  We leave
the nature of such an instability
and its applicability to real graphene as
open questions \cite{herbut_note}.

\section{\label{cnclsn}conclusion}
In this work we have investigated excitonic effects for graphene with
Coulomb interactions, as modeled by massless Dirac fermions.
We have shown that there is power law behavior in a general 4-leg
vertex function in the particle-hole channel. The exponent becomes
complex as the coupling constant $\beta$ is increased above a
critical value. This is analogous to what happens in the problem
of a single MDF interacting with a charged impurity. This non-analytic
behavior can be canceled away for certain combinations of the
vertex function, and we find in particular that it is absent
in the density-density response function.  It is however
retained in a sublattice antisymmetric response function.
Although the power law behavior originates due to short length
scale physics (close approaches of particle-hole pairs), it impacts
the physics at large distances because of the absence of a length
scale in the Hamiltonian.  For finite size systems the transition
appears to be one with broken chiral symmetry, inducing a gap in
the spectrum; however, this interpretation breaks down in the
thermodynamic limit.  We speculate that in this case the transition involves
the formation of a mass gap which fluctuates among different
possible forms, and is a precursor to a true broken symmetry
state which emerges at still larger values of the coupling $\beta$.

We have also calculated the density response in the ladder approximation
numerically using a simplified model Hamiltonian
that occurs in the context of topological insulators,
which has only one
Dirac point and a square Brillouin zone. The calculation was carried out
for both undoped and doped cases.  In the latter case we compared results
for RPA screened Coulomb interactions
and contact interactions in the rungs of the ladders.
While we expect
the former interaction to be more realistic,
both interactions in many respects give similar results.
For Coulomb interactions, we find a strongly enhanced
cusp in the irreducible polarizability at $2k_F$, which
leads to much stronger Friedel oscillations than
expected from the RPA.  We also find a hump-like
structure at stronger interaction scales just below $2k_F$.
For contact interactions
this hump evolves into poles with increasing interaction strength,
whereas no pole is seen for RPA screened Coulomb interactions
in the range of $\beta$ we have studied.  We presented
evidence that the extra structure
is peculiar to our model Hamiltonian, suggesting it may
be present at the surface of a topological insulator.
Finally, we note that with contact interactions, MDF's
do contain a pole at larger $q$
for any positive $u_0$, suggesting a short
wavelength instability.

\acknowledgments
This work was supported by the Binational Science Foundation through
Grant No. 2008256 (HAF and JW), by the NSF through Grant Nos. DMR-1005035 (HAF)
and DMR-0703992 (GM), and by the MEC-Spain via Grant No. FIS2009-08744 (LB).
Numerical calculations described here were performed
on Indiana University's computer cluster Quarry.

\begin{appendix}
\section{}
In this Appendix, we discuss some details of the TLA method.
This may be viewed as a two-dimensional version of the ``tetrahedron
method'' \cite{jepsen_1971,lehmann_1972} which is widely applied for Brillouin zone
integrations of three dimensional systems.
As
mentioned in Ref.~\onlinecite{second-TLA}, when using linear interpolation,
the surfaces $e=\varepsilon(x,y)$
(in this Appendix $x\equiv k_x$ and $y\equiv k_y$) are straight lines
and the parameterizations are easily given
(with $e$ being one of the parameters). There are two cases:
$e<\e_2$ (but larger than $\e_1$) and $e>\e_2$ (but smaller than $\e_3$).
The parameterizations are given by Eqs.~(14) and (16) in Ref.~\onlinecite{second-TLA},
respectively. For convenience we reproduce them here.
\begin{align}
\vec{k}=&\vk_3+\frac{e-\e_3}{\e_3-\e_1}(\vk_3-\vk_1)\notag \\
&+u\left[\frac{e-\e_3}{\e_3-\e_2}(\vk_3-\vk_2)-\frac{e-\e_3}{\e_3-\e_1}(\vk_3-\vk_1)\right] \label{eq-e-gt-e2}
\end{align}
for $e>\e_2$, and
\begin{align}
\vk=&\vk_1+\frac{e-\e_1}{\e_3-\e_1}(\vk_3-\vk_1) \notag \\
&+u\left[\frac{e-\e_1}{\e_2-\e_1}(\vk_2-\vk_1)-\frac{e-\e_1}{\e_3-\e_1}(\vk_3-\vk_1)\right] \label{eq-e-lt-e2}
\end{align}
for $e<\e_2$; in both cases $0\leqslant u \leqslant 1$.

\begin{figure}
\begin{center}
\subfigure[]{\label{fig-gt-e2}\includegraphics[scale=0.4]{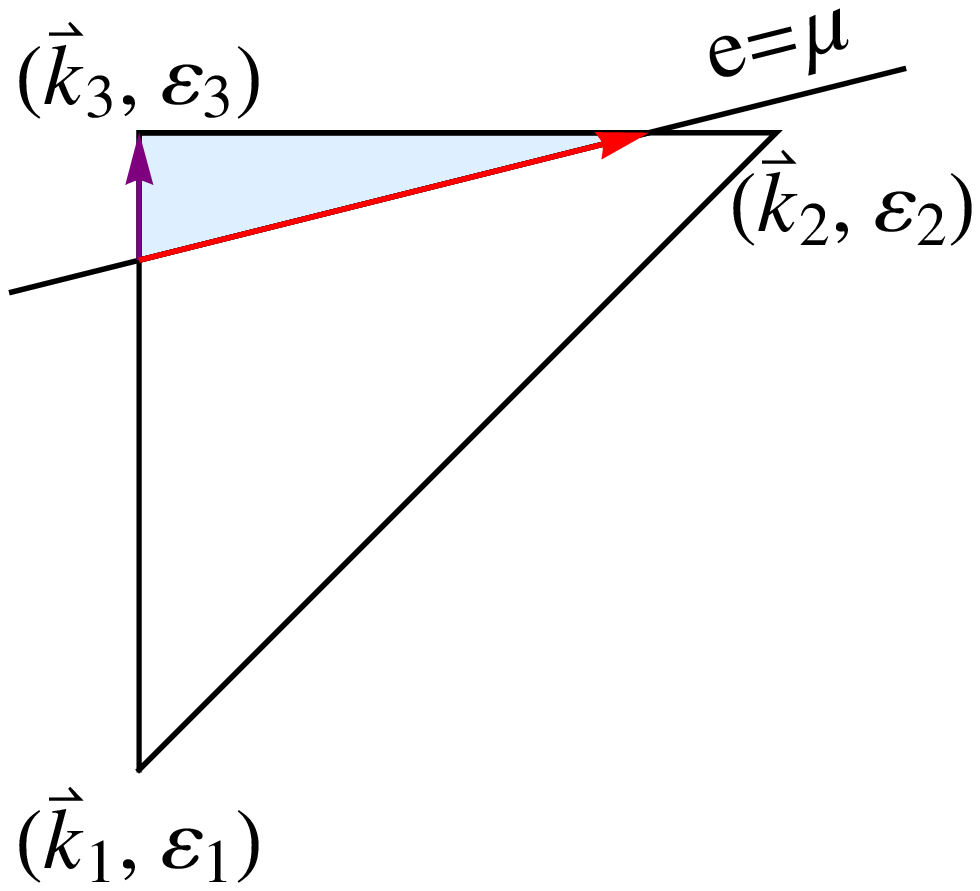}}
\subfigure[]{\label{fig-lt-e2}\includegraphics[scale=0.4]{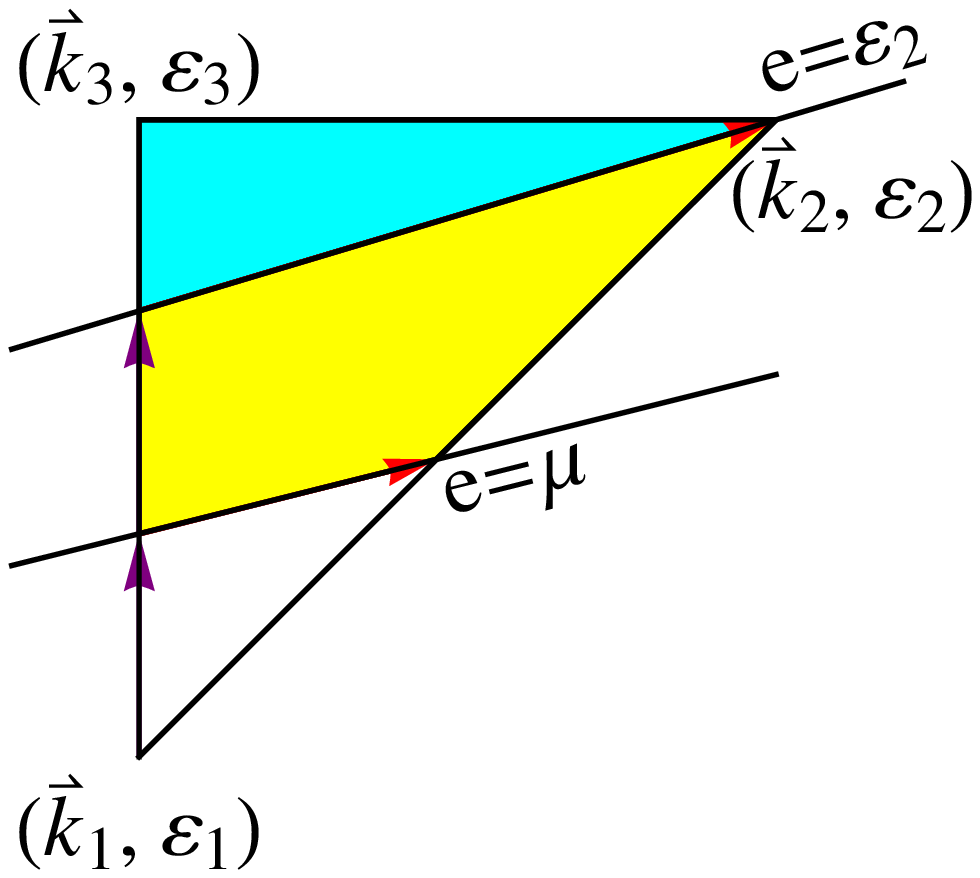}}
\end{center}
\caption{ (Color online) Illustrations for (a) case (ii) and (b)  case (iii). In both subfigures the red and purple arrows label the directions of increasing $e$ and $u$ respectively. However, while they correspond to Eq.~(\ref{eq-e-gt-e2}) in (a), in (b) they correspond to Eq.~(\ref{eq-e-lt-e2}).}
\end{figure}

Our goal is to express the integral over a basic triangle in terms of the values of $\varepsilon(x,y)$ and the integrand at the three corners of the triangle, i.e.
\begin{equation}
I(\mu)=\int_{\bigtriangleup}d^2k\theta(\varepsilon_{\vec{k}}-\mu)f(\vec{k})\approx \sum_{i=1}^3 w_i(\e_1,\e_2,\e_3,\mu)f(\vec{k}_i),
\end{equation}
where $\int_{\triangle}$ means integration over the triangle, $w_i,\,i=1,2,3$ are the weights at the three corners (remember that the corners are labeled so that $\e_1<\e_2<\e_3$). We will also use the shorthand $f_i\equiv f(\vec{k}_i)$ below.

There are four possibilities regarding the value of $\mu$ as compared to $\e_1,\e_2,\e_3$:

(i) $\mu \geqslant \e_3$:

This is trivial and the result is $w_i=0,\,i=1,2,3$.

(ii) $\e_2 \leqslant \mu<\e_3$:

In this case, we need to use Eq.~(\ref{eq-e-gt-e2}) for the parameterization, and
\begin{widetext}
\begin{equation}
I(\mu)\approx \int 'd^2kf(\vec{k}(e,u))=\int_{\mu}^{\e_3}de\int_0^1du\frac{2A(\e_3-e)}{(\e_3-\e_2)(\e_3-\e_1)}f(\vec{k}(e,u)) \equiv I_{\text{I}}(\mu),
\end{equation}
where $\int '$ means integration over the uppermost (light blue)
triangle in Fig.~\ref{fig-gt-e2}, and the extra factor in the integrand is just the Jacobian, with $A$ being the area of the triangle 123 (not the shaded triangle).

We intepolate $f$ linearly, i.e.
\begin{equation}
f(x,y) \approx p_1+p_2 x+p_3y\equiv \sum_{i=1}^3 p_i g_i(x,y),
\end{equation}
where $g_1\equiv 1$, $g_2\equiv x$, $g_3\equiv y$, and the coefficients $p_i,\,i=1,2,3$ are determined by solving the equations $\sum_{i=1}^3 p_ig_i(x_j,y_j)=f_j$. The result is
\begin{equation}
p_1 =\frac{(x_3y_2-x_2y_3)f_1+(x_1y_3-x_3y_1)f_2+(x_2y_1-x_1y_2)f_3}{x_2y_1-x_3y_1-x_1y_2+x_3y_2+x_1y_3-x_2y_3},
\end{equation}
$p_2$ and $p_3$ have the same denominator but with the numerator being $[(y_3-y_2)f_1+(y_1-y_3)f_2+(y_2-y_1)f_3]$ for $p_2$ and $[(x_2-x_3)f_1+(x_3-x_1)f_2+(x_1-x_2)f_3]$ for $p_3$.
\begin{equation}
I_{\text{I}}(\mu) \approx \sum_{i=1}^3 A p_i \int_{\mu}^{\e_3}de\int_0^1du \frac{2(\e_3-e)}{(\e_3-\e_2)(\e_3-\e_1)}g_i \equiv \sum_{i=1}^3 A p_i I_{\text{I}}(\mu)_i.
\end{equation}
The results of the integrations are

\begin{align}
I_{\text{I}}(\mu)_1&=\int_{\mu}^{\e_3}de\frac{2(\e_3-e)}{(\e_3-\e_2)(\e_3-\e_1)}\int _0^1du=\frac{(\e_3-\mu)^2}{(\e_3-\e_2)(\e_3-\e_1)},\\
I_{\text{I}}(\mu)_2&=\int_{\mu}^{\e_3}de\frac{2(\e_3-e)}{(\e_3-\e_2)(\e_3-\e_1)}\int _0^1du \left\{x_3+\frac{e-\e_3}{\e_3-\e_1}(x_3-x_1)+u\left[\frac{e-\e_3}{\e_3-\e_2}(x_3-x_2)-\frac{e-\e_3}{\e_3-\e_1}(x_3-x_1)\right]\right\} \notag \\
&=\frac{(\e_3-\mu)^2}{(\e_3-\e_2)(\e_3-\e_1)}\left[x_3-\frac{1}{3}(\e_3-\mu)\left(\frac{x_3-x_1}{\e_3-\e_1}+\frac{x_3-x_2}{\e_3-\e_2}\right)\right],
\end{align}
and $I_{\text{I}}(\mu)_3$ is the same as $I_{\text{I}}(\mu)_2$ but with the $x$'s replaced by $y$'s. %these are jv1,jv2,jv3 in my notes.
Finally, we have
\begin{equation}
w_1 =\frac{A\left[(x_3y_2-x_2y_3)I_{\text{I}}(\mu)_1+(y_3-y_2)I_{\text{I}}(\mu)_2+(x_2-x_3)I_{\text{I}}(\mu)_3\right]}{x_2y_1-x_3y_1-x_1y_2+x_3y_2+x_1y_3-x_2y_3},
\end{equation}
$w_2$ and $w_3$ have the same denominator but with the numerator being $A[(x_1y_3-x_3y_1)I_{\text{I}}(\mu)_1+(y_1-y_3)I_{\text{I}}(\mu)_2+(x_3-x_1)I_{\text{I}}(\mu)_3]$ for $w_2$ and $A[(x_2y_1-x_1y_2)I_{\text{I}}(\mu)_1+(y_2-y_1)I_{\text{I}}(\mu)_2+(x_1-x_2)I_{\text{I}}(\mu)_3]$ for $w_3$.

(iii)$\e_1 <\mu <\e_2$

In this case, $I(\mu)\approx \int ''d^2kf(\vec{k}(e,u))$ where the integration is over the uppermost (cyan)
triangle and the yellow quadrilateral in Fig.~\ref{fig-lt-e2}. The former is approximately the same (to linear order of the size of the basic triangles) as $I_{\text{I}}(\e_2)$, while the latter is
\begin{equation}
I_{\text{II}}(\mu)\approx \sum_{i=1}^3 Ap_i \int_{\mu}^{\e_2}de\frac{2(e-\e_1)}{(\e_2-\e_1)(\e_3-\e_1)}\int_0^1du g_i \equiv \sum_{i=1}^3A p_i I_{\text{II}}(\mu)_i.
\end{equation}
The $I_{\text{I}}(\mu)_i$'s in the formula given above for case (ii) should be replaced by
\begin{eqnarray}
I_{\text{I}}(\e_2)_1+I_{\text{II}}(\mu)_1&=&\frac{\e_3-\e_2}{\e_3-\e_1}+\int_{\mu}^{\e_2}de\frac{2(e-\e_1)}{(\e_2-\e_1)(\e_3-\e_1)}\int _0^1du \nonumber \\
&=&\frac{\e_3-\e_2}{\e_3-\e_1}+\frac{(\e_2-\e_1)^2-(\mu-\e_1)^2}{(\e_2-\e_1)(\e_3-\e_1)},\\
I_{\text{I}}(\e_2)_2+I_{\text{II}}(\mu)_2&=&\frac{\e_3-\e_2}{\e_3-\e_1}\left[x_3-\frac{1}{3}(\e_3-\e_2)\left(\frac{x_3-x_1}{\e_3-\e_1}+\frac{x_3-x_2}{\e_3-\e_2}\right)\right]+\int_{\mu}^{\e_2}de\frac{2(e-\e_1)}{(\e_2-\e_1)(\e_3-\e_1)} \nonumber \\
  &&\times \int _0^1du \left\{x_1+\frac{e-\e_1}{\e_3-\e_1}(x_3-x_1)+u\left[\frac{e-\e_1}{\e_2-\e_1}(x_2-x_1)-\frac{e-\e_1}{\e_3-\e_1}(x_3-x_1)\right]\right\} \nonumber\\
&=&\frac{\e_3-\e_2}{\e_3-\e_1}\left[x_3-\frac{1}{3}(\e_3-\e_2)\left(\frac{x_3-x_1}{\e_3-\e_1}+\frac{x_3-x_2}{\e_3-\e_2}\right)\right]+\frac{1}{(\e_2-\e_1)(\e_3-\e_1)} \nonumber \\
%&&-\frac{(\e_2-\mu)}{3(\e_1-\e_2)^2(\e_1-\e_3)^2} \left\{\e_3 \mu ^2\left(-x_1+x_2\right)+\e _2^2\left[\e _3 \left(-4x_1+x_2\right)+\mu \left(x_1-x_3\right)\right]\right.\nonumber \\
%&&+\e_2 \mu \left[\e_3 \left(-4x_1+x_2\right)+\mu \left(x_1-x_3\right)\right]+\e_2^3\left(x_1-x_3\right)\right. \nonumber \\
%&& \left. +3\e_1^3\left(2x_1-x_2+x_3\right)+\e_1\left[6\e_3\left(2\e_2+m)x_1-\right]\right\},
&&\times \left\{\left[(\e_2-\e_1)^2-(\mu-\e_1)^2\right]x_1+\frac{1}{3}\left[(\e_2-\e_1)^3-(\mu-\e_1)^3\right]\left(\frac{x_2-x_1}{\e_2-\e_1}+\frac{x_3-x_1}{\e_3-\e_1}\right)\right\}
\end{eqnarray}
and $I_{\text{II}}(\e_2)_3+I_{\text{II}}(\mu)_3$, which is the same as $I_{\text{I}}(\e_2)_2+I_{\text{II}}(\mu)_2$ but with the $x$'s replaced by $y$'s.

(iv)$\mu \leqslant \e_1$

This case is trivial too, we simply have $w_1=w_2=w_3=1/3$.
\end{widetext}
\end{appendix}

%%%%%%%%%%%%%%%%%%%%%%%%%%%%%%%%%%%%%%%%%%%%%%
%\bibliographystyle{apsrev}


\begin{thebibliography}{36}
\expandafter\ifx\csname natexlab\endcsname\relax\def\natexlab#1{#1}\fi
\expandafter\ifx\csname bibnamefont\endcsname\relax
  \def\bibnamefont#1{#1}\fi
\expandafter\ifx\csname bibfnamefont\endcsname\relax
  \def\bibfnamefont#1{#1}\fi
\expandafter\ifx\csname citenamefont\endcsname\relax
  \def\citenamefont#1{#1}\fi
\expandafter\ifx\csname url\endcsname\relax
  \def\url#1{\texttt{#1}}\fi
\expandafter\ifx\csname urlprefix\endcsname\relax\def\urlprefix{URL }\fi
\providecommand{\bibinfo}[2]{#2}
\providecommand{\eprint}[2][]{\url{#2}}

\bibitem[{\citenamefont{Neto et~al.}(2009)\citenamefont{Neto, F.Guinea,
  N.M.R.Peres, K.S.Novoselov, and A.K.Geim}}]{Castro_Neto_RMP}
\bibinfo{author}{\bibfnamefont{A.~C.} \bibnamefont{Neto}},
  \bibinfo{author}{\bibnamefont{F.Guinea}},
  \bibinfo{author}{\bibnamefont{N.M.R.Peres}},
  \bibinfo{author}{\bibnamefont{K.S.Novoselov}}, \bibnamefont{and}
  \bibinfo{author}{\bibnamefont{A.K.Geim}}, \bibinfo{journal}{Rev.\ Mod.\
  Phys.} \textbf{\bibinfo{volume}{81}}, \bibinfo{pages}{109}
  (\bibinfo{year}{2009}).

\bibitem[{Das()}]{DasSarma2010}
\bibinfo{note}{S. Das Sarma, S. Adam, E.H. Hwang and E. Rossi,
  arXiv:1003.4731}.

\bibitem[{\citenamefont{Gonzalez et~al.}(1994)\citenamefont{Gonzalez, F.Guinea,
  and Vozmediano}}]{Gonzalez_1994}
\bibinfo{author}{\bibfnamefont{J.}~\bibnamefont{Gonzalez}},
  \bibinfo{author}{\bibnamefont{F.Guinea}}, \bibnamefont{and}
  \bibinfo{author}{\bibfnamefont{M.}~\bibnamefont{Vozmediano}},
  \bibinfo{journal}{Nuc.\ Phys.\ B} \textbf{\bibinfo{volume}{424}},
  \bibinfo{pages}{595} (\bibinfo{year}{1994}).

\bibitem[{dru()}]{drut123}
\bibinfo{note}{J. E. Drut and T. A. L\"{a}hde, Phys. Rev. Lett. {\bf 102},
  026802 (2009); Phys. Rev. B {\bf 79}, 165425 (2009); {\it ibid.} {\bf 79},
  241405 (2009)}.

\bibitem[{shy()}]{shytov2007}
\bibinfo{note}{A. V. Shytov and M. I. Katsnelson and L. S. Levitov, Phys. Rev.
  Lett. {\bf 99}, 236801 (2007)}.

\bibitem[{per({\natexlab{a}})}]{pereira166802}
\bibinfo{note}{V. M. Pereira and J. Nilsson and A. H. Castro Neto, Phys. Rev.
  Lett. {\bf 99}, 166802 (2007)}.

\bibitem[{bis()}]{biswas205122}
\bibinfo{note}{R. R. Biswas and S. Sachdev and D. T. Son, Phys. Rev. B {\bf
  76}, 205122 (2007)}.

\bibitem[{ter()}]{terekhov076803}
\bibinfo{note}{I. S. Terekhov and A. I. Milstein and V. N. Kotov and O. P.
  Sushkov, Phys. Rev. Lett. {\bf 100}, 076803 (2008)}.

\bibitem[{per({\natexlab{b}})}]{pereira085101}
\bibinfo{note}{V. M. Pereira and V. N. Kotov and A. H. Castro Neto, Phys. Rev.
  B {\bf 78}, 085101 (2008)}.

\bibitem[{kot()}]{kotov075433}
\bibinfo{note}{V. N. Kotov and V. M. Pereira and B. Uchoa, Phys. Rev. B {\bf
  78}, 075433 (2008)}.

\bibitem[{\citenamefont{Schwabl}(2008)}]{schwabl_book}
\bibinfo{author}{\bibfnamefont{F.}~\bibnamefont{Schwabl}}, in
  \emph{\bibinfo{booktitle}{Advanced Quantum Mechanics}}
  (\bibinfo{publisher}{Springer-Verlag}, \bibinfo{address}{Heidelberg},
  \bibinfo{year}{2008}).

\bibitem[{fet()}]{fetter}
\bibinfo{note}{A. L. Fetter and J. D. Walecka, {\it Quantum Theory of
  Many-Particle Systems}, Chap. 4, \S11 (Dover, 2003)}.

\bibitem[{\citenamefont{Nelson}(2002)}]{nelson_2002}
\bibinfo{author}{\bibfnamefont{D.}~\bibnamefont{Nelson}}, in
  \emph{\bibinfo{booktitle}{Defects and Geometry in Condensed Matter Physics}}
  (\bibinfo{publisher}{Cambridge University Press}, \bibinfo{address}{New
  York}, \bibinfo{year}{2002}).

\bibitem[{\citenamefont{Wang et~al.}(2010)\citenamefont{Wang, Fertig, and
  Murthy}}]{asym}
\bibinfo{author}{\bibfnamefont{J.}~\bibnamefont{Wang}},
  \bibinfo{author}{\bibfnamefont{H.~A.} \bibnamefont{Fertig}},
  \bibnamefont{and} \bibinfo{author}{\bibfnamefont{G.}~\bibnamefont{Murthy}},
  \bibinfo{journal}{Phys. Rev. Lett.} \textbf{\bibinfo{volume}{104}},
  \bibinfo{pages}{186401} (\bibinfo{year}{2010}).

\bibitem[{\citenamefont{Cheianov and Falko}(2006)}]{cheianov_2006}
\bibinfo{author}{\bibfnamefont{V.}~\bibnamefont{Cheianov}} \bibnamefont{and}
  \bibinfo{author}{\bibfnamefont{V.}~\bibnamefont{Falko}},
  \bibinfo{journal}{Phys.\ Rev.\ Lett.} \textbf{\bibinfo{volume}{97}},
  \bibinfo{pages}{226801} (\bibinfo{year}{2006}).

\bibitem[{\citenamefont{Brey et~al.}(2007)\citenamefont{Brey, Fertig, and {Das
  Sarma}}}]{Brey_2007b}
\bibinfo{author}{\bibfnamefont{L.}~\bibnamefont{Brey}},
  \bibinfo{author}{\bibfnamefont{H.}~\bibnamefont{Fertig}}, \bibnamefont{and}
  \bibinfo{author}{\bibfnamefont{S.}~\bibnamefont{{Das Sarma}}},
  \bibinfo{journal}{Phys.\ Rev.\ Lett.} \textbf{\bibinfo{volume}{99}},
  \bibinfo{pages}{116802} (\bibinfo{year}{2007}).

\bibitem[{\citenamefont{Wehling et~al.}(2007)\citenamefont{Wehling, Balatsky,
  Katsnelson, Lichtenstein, Scharnberg, and Wiesendanger}}]{wehling_2007}
\bibinfo{author}{\bibfnamefont{T.}~\bibnamefont{Wehling}},
  \bibinfo{author}{\bibfnamefont{A.~V.} \bibnamefont{Balatsky}},
  \bibinfo{author}{\bibfnamefont{M.~I.} \bibnamefont{Katsnelson}},
  \bibinfo{author}{\bibfnamefont{A.~I.} \bibnamefont{Lichtenstein}},
  \bibinfo{author}{\bibfnamefont{K.}~\bibnamefont{Scharnberg}},
  \bibnamefont{and}
  \bibinfo{author}{\bibfnamefont{R.}~\bibnamefont{Wiesendanger}},
  \bibinfo{journal}{Phys. Rev. B} \textbf{\bibinfo{volume}{75}},
  \bibinfo{pages}{125425} (\bibinfo{year}{2007}).

\bibitem[{\citenamefont{Yazyev and Helm}(2007)}]{yazyev_2007}
\bibinfo{author}{\bibfnamefont{O.}~\bibnamefont{Yazyev}} \bibnamefont{and}
  \bibinfo{author}{\bibfnamefont{L.}~\bibnamefont{Helm}},
  \bibinfo{journal}{Phys.\ Rev.\ B} \textbf{\bibinfo{volume}{75}},
  \bibinfo{pages}{125408} (\bibinfo{year}{2007}).

\bibitem[{\citenamefont{Khveshchenko}(2001)}]{Khveshchenko}
\bibinfo{author}{\bibfnamefont{D.~V.} \bibnamefont{Khveshchenko}},
  \bibinfo{journal}{Phys. Rev. Lett.} \textbf{\bibinfo{volume}{87}},
  \bibinfo{eid}{246802} (\bibinfo{year}{2001}).

\bibitem[{Gor()}]{Gorbar2002}
\bibinfo{note}{E. V. Gorbar, V. P. Gusynin, V. A. Miransky, and I. A. Shovkovy,
  Phys. Rev. B {\bf 66}, 045108 (2002)}.

\bibitem[{\citenamefont{Merzbacher}(1970)}]{merzbacher1970}
\bibinfo{author}{\bibfnamefont{E.}~\bibnamefont{Merzbacher}},
  \emph{\bibinfo{title}{Quantum Mechanics, 2nd Edition}}
  (\bibinfo{publisher}{Wiley}, \bibinfo{address}{New York},
  \bibinfo{year}{1970}).

\bibitem[{Num()}]{NumericalRecipe}
\bibinfo{note}{W.~H. Press {\it et al.}, {\it Numerical Recipes, 3rd Edition}
  (Cambridge University Press, 2007)}.

\bibitem[{ref()}]{ref_credit}
\bibinfo{note}{The authors thank the anonymous referee of Ref.
  \onlinecite{asym} for pointing this out.}

\bibitem[{\citenamefont{M.~Hazewinkel}(2001)}]{hazewinkel_2001}
\bibinfo{author}{\bibfnamefont{E.}~\bibnamefont{M.~Hazewinkel}},
  \emph{\bibinfo{title}{The Wiener-Hopf Method}}
  (\bibinfo{publisher}{Springer}, \bibinfo{year}{2001}).

\bibitem[{\citenamefont{Qi et~al.}(2006)\citenamefont{Qi, Wu, and
  Zhang}}]{Qi2004}
\bibinfo{author}{\bibfnamefont{X.-L.} \bibnamefont{Qi}},
  \bibinfo{author}{\bibfnamefont{Y.-S.} \bibnamefont{Wu}}, \bibnamefont{and}
  \bibinfo{author}{\bibfnamefont{S.-C.} \bibnamefont{Zhang}},
  \bibinfo{journal}{Phys. Rev. B} \textbf{\bibinfo{volume}{74}},
  \bibinfo{pages}{045125} (\bibinfo{year}{2006}).

\bibitem[{\citenamefont{Qi et~al.}(2008)\citenamefont{Qi, Hughes, and
  Zhang}}]{Qi2008}
\bibinfo{author}{\bibfnamefont{X.-L.} \bibnamefont{Qi}},
  \bibinfo{author}{\bibfnamefont{T.~L.} \bibnamefont{Hughes}},
  \bibnamefont{and} \bibinfo{author}{\bibfnamefont{S.-C.} \bibnamefont{Zhang}},
  \bibinfo{journal}{Phys. Rev. B} \textbf{\bibinfo{volume}{78}},
  \bibinfo{pages}{195424} (\bibinfo{year}{2008}).

\bibitem[{fir()}]{first-TLA}
\bibinfo{note}{J. A. Ashraff and P. D. Loly, J. Phys. C {\bf 20}, 4823 (1987)}.

\bibitem[{sec()}]{second-TLA}
\bibinfo{note}{G. Wiesenekker, G. te Velde and E. J. Baerends, J. Phys. C {\bf
  21}, 4263 (1988)}.

\bibitem[{tri()}]{triangles_fixed}
\bibinfo{note}{The formulas provided in Ref. \onlinecite{second-TLA} require
  some modification for this application. When a triangle is completely
  ``occupied'' (i.e., all three points inside the Fermi surface), the weights
  are still dependent on $\varepsilon_{\vec{k}_i}$ or
  $\varepsilon_{\vec{k}_i+\vec{q}}$ [for $\theta(\varepsilon_{\vec{k}}-\mu)$
  and $\theta(\varepsilon_{\vec{k}+\vec{q}}-\mu)$ terms respectively], where
  $i=1,2,3$ labels the corners. If there are grid points where
  $\varepsilon_{\vec{k}}=\varepsilon_{\vec{k}+\vec{q}}$ and the weights
  calculated using $\varepsilon_i=\varepsilon_{\vec{k}_i}$ and
  $\varepsilon_i=\varepsilon_{\vec{k}_i+\vec{q}}$ are different (which in
  general is the case), then the first line in Eq.~(\ref{eq-K-mu-gt-0}) is
  infinite. To avoid this situation, we set the weights at the 3 corners to 1/3
  of the area of a triangle if the triangle is fully occupied, i.e.
  $\varepsilon_i \leq \mu$, $i=1,2,3$. After calculating the weights at every
  grid point, we divide each small square into triangles along the other
  diagonal and re-calculate the weights. The final weights are the averages of
  the results of the two calculations. See the appendix for more detail.}

\bibitem[{\citenamefont{Hwang and Das~Sarma}(2007)}]{Hwang-polarizability}
\bibinfo{author}{\bibfnamefont{E.~H.} \bibnamefont{Hwang}} \bibnamefont{and}
  \bibinfo{author}{\bibfnamefont{S.}~\bibnamefont{Das~Sarma}},
  \bibinfo{journal}{Phys. Rev. B} \textbf{\bibinfo{volume}{75}},
  \bibinfo{pages}{205418} (\bibinfo{year}{2007}).

\bibitem[{\citenamefont{Wunsch et~al.}(2006)\citenamefont{Wunsch, Stauber,
  Sols, and Guinea}}]{wunsch_2006}
\bibinfo{author}{\bibfnamefont{B.}~\bibnamefont{Wunsch}},
  \bibinfo{author}{\bibfnamefont{T.}~\bibnamefont{Stauber}},
  \bibinfo{author}{\bibfnamefont{F.}~\bibnamefont{Sols}}, \bibnamefont{and}
  \bibinfo{author}{\bibfnamefont{F.}~\bibnamefont{Guinea}},
  \bibinfo{journal}{New J. Phys.} \textbf{\bibinfo{volume}{8}},
  \bibinfo{pages}{318} (\bibinfo{year}{2006}).

\bibitem[{\citenamefont{Brey and Halperin}(1989)}]{brey_1989}
\bibinfo{author}{\bibfnamefont{L.}~\bibnamefont{Brey}} \bibnamefont{and}
  \bibinfo{author}{\bibfnamefont{B.}~\bibnamefont{Halperin}},
  \bibinfo{journal}{Phys. Rev. B.} \textbf{\bibinfo{volume}{40}},
  \bibinfo{pages}{11634} (\bibinfo{year}{1989}).

\bibitem[{fri()}]{friedel}
\bibinfo{note}{See, for example, Ref. \onlinecite{Brey_2007b} and references
  therein.}

\bibitem[{her()}]{herbut_note}
\bibinfo{note}{See, for example, I.F. Herbut, V. Juricic, and B. Roy, Phys.
  Rev. B {\bf 79}, 085116 (2009) and references therein.}

\bibitem[{\citenamefont{Jepsen and Andersen}(1971)}]{jepsen_1971}
\bibinfo{author}{\bibfnamefont{O.}~\bibnamefont{Jepsen}} \bibnamefont{and}
  \bibinfo{author}{\bibfnamefont{O.}~\bibnamefont{Andersen}},
  \bibinfo{journal}{Solid State Commun.} \textbf{\bibinfo{volume}{9}},
  \bibinfo{pages}{1763} (\bibinfo{year}{1971}).

\bibitem[{\citenamefont{Lehmann and Taut}(1972)}]{lehmann_1972}
\bibinfo{author}{\bibfnamefont{G.}~\bibnamefont{Lehmann}} \bibnamefont{and}
  \bibinfo{author}{\bibfnamefont{M.}~\bibnamefont{Taut}},
  \bibinfo{journal}{Phys. Status Solidi B} \textbf{\bibinfo{volume}{54}}
  (\bibinfo{year}{1972}).

\end{thebibliography}
\end{document}